\documentclass{cernrep} 
\usepackage{texnames}
\usepackage[T1]{fontenc}
 \usepackage{bbding} 
 \usepackage{color}

\newcommand{\eq}[1]{Eq.~(\ref{#1})}

\newcommand{\nn}{\nonumber}

\newcommand{\Dslash}{\!\not\!\!  D}
\newcommand{\be}{\begin{equation}}
\newcommand{\ee}{\end{equation}}
\newcommand{\bea}{\begin{eqnarray}}
\newcommand{\eea}{\end{eqnarray}}
\newcommand{\bc}{\begin{center}}
\newcommand{\ec}{\end{center}}

\def\stw{s_{\theta_W}}
\def\ctw{c_{\theta_W}}
\def\ttw{t_{\theta_W}}

\begin{document}
\title{Higgs Physics}
 
\author{Alex Pomarol}

\institute{Dept. de F\'isica, Universitat Aut{\`o}noma de Barcelona, 08193~Bellaterra,~Barcelona}

\maketitle 

\begin{abstract}
With the discovery of the Higgs,  we  have  access to a plethora of new physical processes that  allow us  to further test the SM and beyond.  We show a convenient way to parametrize  these physics using  an effective theory for Higgs couplings,
discussing the importance of the basis selection,  predictions from  a SM effective field theory,
and  possible ways to measure these couplings with special attention to the high-energy regime. 
Predictions from the  MSSM and MCHM, with the  comparison with data,  are also provided.
\end{abstract}
 
\section{Motivation}

\begin{figure}
\centering\includegraphics[width=.6\linewidth]{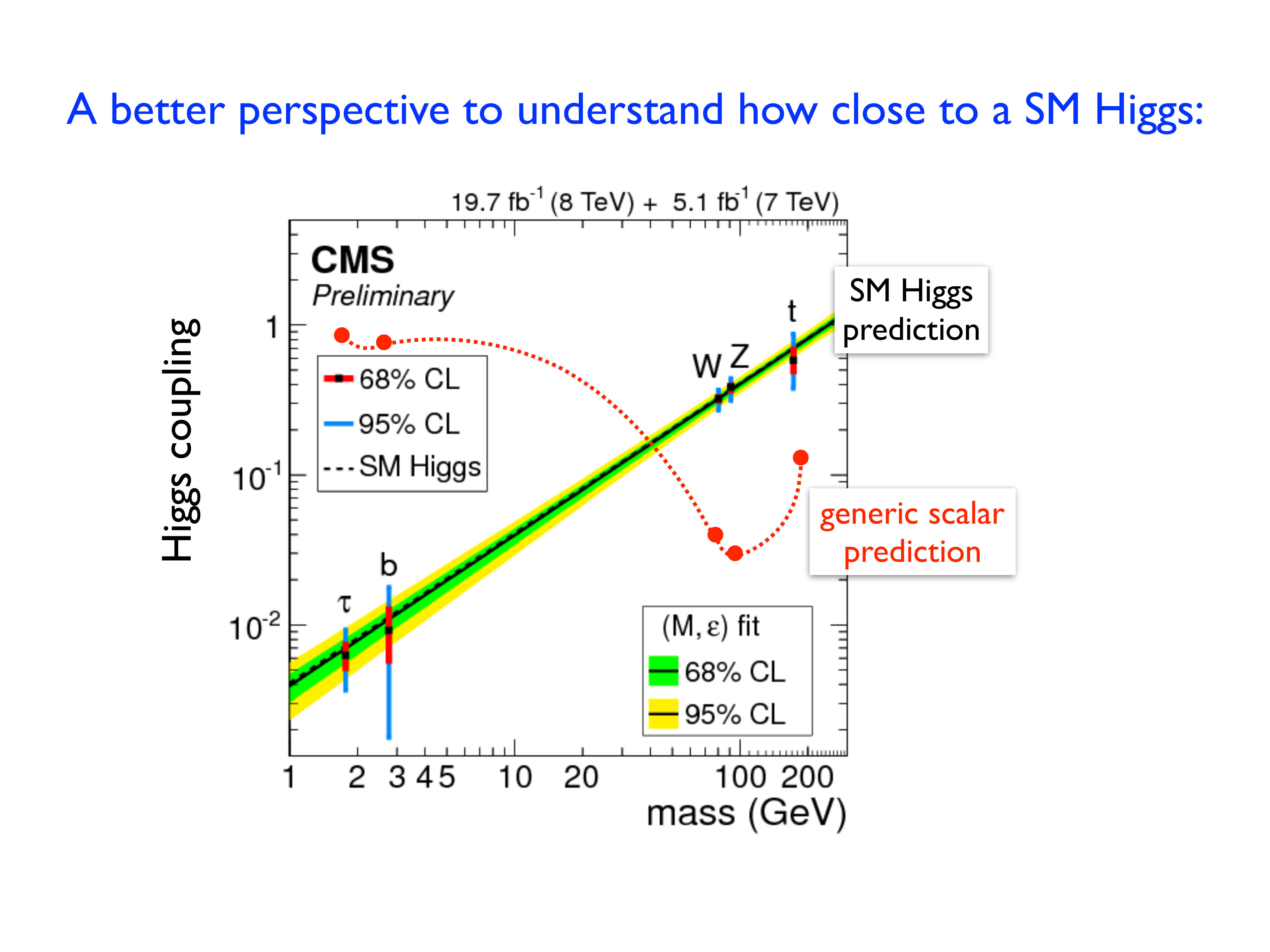}
\caption{Fit of the Higgs couplings, $g_{ff}^h$ and $\sqrt{g_{VV}^h/2v}$,  and predictions from the SM  \cite{CMS:2014}.
A generic scalar  would have couplings to the SM particles  laying in any point of this plane, as the example   shown in red.
The experimental data clearly favors a SM Higgs.}
\label{Higgscouplings}
\end{figure}

The 4th of July of 2012  marked a milestone in particle physics, as CERN announced the discovery of a new particle
whose properties  were in accordance with the sought-after  Higgs boson \cite{Higgsdisc}.
Since then,  we have been accumulating more  and more data and  measuring   more decay channels, 
 increasing the  significance  of the   discovery while   keeping at the same time a good agreement with the predictions from the Standard Model (SM) Higgs  \cite{CMS:2014,ATLAS14}. To   appreciate this agreement,  it is 
convenient to plot the experimental fit to Higgs couplings in the coupling--mass plane, 
as shown in Fig.~\ref{Higgscouplings}  by courtesy of CMS \cite{CMS:2014}. 
Were this  new particle not the SM Higgs, we would have expected its couplings to
 lay  on any point of this plane,  and therefore differing significantly  from   the SM predictions. 
 As an example, let us consider a scalar coming from a weak-doublet  
 not being (the main) responsible for electroweak symmetry breaking (EWSB). 
 This scalar  could    have couplings to fermions as large as $O(1)$, but very small couplings to $Z/W$.
 These  predictions are shown in   red  in Fig.~\ref{Higgscouplings}.
 Data clearly disfavours  this type of scalars as compared with the  SM Higgs whose predictions lay on a straight line.
We can then say today that the SM Higgs is significantly supported by the experimental data, leaving most competitors far behind.

Having discovered the Higgs,  we  have now experimental  access to  new processes 
 that  will help  us  to  test the SM and beyond. 
 There is a fundamental  aspect that makes   Higgs physics  very special:
the Higgs  is the only particle of the SM that its lightness ($m_h\sim 125$ GeV $\ll$ $M_P$) is not expected on theoretical grounds, requiring the presence of new physics beyond the SM (BSM) at the TeV. This is referred as  the hierarchy problem.
This makes the Higgs boson one of the most sensitive SM particle to BSM effects, and therefore the measurement of its properties  one of  the  best ways to  indirectly discover  new physics and help to discriminate between different BSMs. As an example,  two of the most well-motivated BSM scenarios, the minimal supersymmetric SM and
the  composite Higgs, predict, as we will see below,   sizeable corrections to the    Higgs couplings.
In few words,  natural theories  explaining the lightness of the Higgs  demand  the Higgs to
be  SM-like only in a first approximation, predicting  departures from the SM predictions to be seen in the near future.

\section{Effective Higgs couplings}

To characterize the most  interesting Higgs processes, it is convenient to 
parametrize, in  the most  general way possible,   the couplings of the Higgs to the SM particles.
For this purpose we will write an  effective theory for the Higgs couplings,  ${\cal L}_h$.
We will define  ${\cal L}_h$  in position-space, as it   makes it simpler to eliminate redundancies.
Our only  approximation at this point will be to assume
that the momenta $q$ in  the Higgs form-factors are smaller than
a heavy scale $\Lambda$  associated with the BSM physical scale, $q/\Lambda\ll1$. 
This is equivalent to say   that we can make an expansion in derivatives $D_\mu/\Lambda$ in ${\cal L}_h$. 
 We leave for later the implications when  an expansion of  SM fields over  $\Lambda$ can be also carried out.
We assume  that the interactions preserve SU(3)$_c$$\times$U(1)$_{\rm EM}$, with the Higgs defined as 
a neutral CP-even scalar field.

We split  the Higgs couplings  in two sets. One set that consists of what  we call {\it primary} Higgs couplings
and the other set containing the rest. 
These primaries, as we will explain later, play an important role,  both theoretically and phenomenologically.
We then write
\be
{\cal L}_h={\cal L}_h^{\rm primary}+\Delta {\cal L}_h\, .
\ee
We will only keep interactions up to order $O(h^3)$, $O(h\partial^2V^2)$ and $O(hVf^2)$ since they are the most relevant for Higgs phenomenology (adding more derivatives will be suppressed by inverse powers of $\Lambda$, and adding more fields makes the interactions harder to be observed at colliders since they will be further suppressed by phase space).  
Then, for CP-conserving couplings,   we have  {\it without loss of  generality}
\footnote{From here and on, all Higgs-coupling coefficients are defined real.}
\bea
{\cal L}_h^{\rm primary}&=&
g^h_{VV}\,  h \left[W^{+\, \mu} W^-_\mu+\frac{1}{2\ctw^2}Z^\mu Z_\mu\right]
+\frac{1}{6}\, g_{3h}\, h^3
+ g^h_{ff}\, \left(h\bar f_Lf_R+h.c.\right)\nn\\ 
&+&\kappa_{GG}\,\frac{h}{2v}G^{A\, \mu\nu}G^A_{\mu\nu}
+\kappa_{\gamma\gamma}\,\frac{h}{2v}
A^{\mu\nu}A_{\mu\nu}
+\kappa_{Z \gamma }\frac{h}{v} A^{\mu\nu} Z_{\mu\nu}\, ,
\label{prim}
 \eea
and 
\bea
\Delta {\cal L}_h&=&
  \delta g^h_{ZZ}\, h \frac{Z^\mu Z_\mu}{2\ctw^2}
+g^h_{Z\!f\!f}\,\frac{h}{2v}\left(Z_\mu J^\mu_{N}+h.c.\right)+g^h_{W\!f\!f'}\,\frac{h}{v}\left(W^+_\mu J^\mu_{C}+h.c.\right)\nn\\
&+&\kappa_{WW}\,\frac{h}{v}
 W^{+\, \mu\nu}W^-_{\mu\nu}
+\kappa_{ZZ}\,\frac{h}{2v} Z^{\mu\nu}Z_{\mu\nu}\, ,
\label{rest}
 \eea
where $J^\mu_{N}=\bar f\gamma^\mu f$ (for $f=f_L,f_R$) and 
$J^\mu_{C}=\bar f\gamma^\mu f'$  are respectively the neutral and charged currents.
Flavour indices are implicit. 
We also defined $\ctw\equiv \cos\theta_W$ where $\theta_W$ is the weak-angle,
and $G^A_{\mu\nu}\equiv \partial_\mu  G^A_\nu-\partial_\nu G^A_\mu$ for gluons, and similarly for the photon, $A_\mu$, the $Z_\mu$ and $W^+_\mu$.
We can use field redefinitions  to
rewrite  the couplings  in \eq{prim} and \eq{rest}  in a different way.
For example,  some linear combinations of the contact-interactions  $hV_\mu J^\mu$ could be written as 
interactions of the type $ hV_\mu \partial_\nu F^{\mu\nu}$ \cite{Giudice:2007fh} by   the redefinition 
 $V_\mu\to (1+\alpha h)V_\mu$, with an appropriate $\alpha$, in the full Lagrangian (and using integration by parts).
Nevertheless, we consider that \eq{prim} and \eq{rest} are the most convenient  way to write the Higgs couplings.  Our parametrization of Higgs couplings gives priority to operators with the largest number of fields (as opposed to
operators with more derivatives), as this is important when estimating the size of the couplings or 
looking for the dominant effects in the high-energy regime, as we will show later.

For CP-violating couplings we have
\bea
{\cal L}_{h}^{\rm primary}&=&  \delta \tilde g_{hff}\, \left(ih\bar f_Lf_R+h.c.\right)
+\widetilde\kappa_{GG}\,\frac{h}{2v}G^{A\, \mu\nu}\widetilde G^A_{\mu\nu}
+\widetilde\kappa_{\gamma\gamma}\,
\frac{h}{2v}A^{\mu\nu}\widetilde A_{\mu\nu}
+\widetilde\kappa_{Z \gamma }\frac{h}{v} A^{\mu\nu}\widetilde Z_{\mu\nu}\, ,\ \ \ \ \ \label{primCPV}\\ 
\Delta{\cal L}_{h}&=&
 \tilde g^h_{Z\!f\!f}\,\frac{h}{2v}\left(iZ_\mu J^\mu_{N}+h.c.\right)+\tilde g^h_{W\!f\!f'}\,\frac{h}{v}\left(iW^+_\mu J^\mu_{C}+h.c.\right)\nn\\
&+&\widetilde \kappa_{WW}\,\frac{h}{v}
 W^{+\, \mu\nu}\widetilde W^-_{\mu\nu}
+\widetilde \kappa_{ZZ}\,\frac{h}{2v} Z^{\mu\nu}\widetilde Z_{\mu\nu}\, ,
\label{restCPV}
\eea
where  $\widetilde G^{A\, \mu\nu}=\epsilon^{\mu\nu\rho\sigma}G^A_{\rho\sigma}/2$ and similarly for other gauge bosons.

It is important to understand the implications of global symmetries in the Higgs couplings.
In particular, if the Higgs couplings are induced from  BSMs that respect a custodial SU(2) symmetry \cite{Weinberg:1979bn} 
only weakly broken by the gauging of  U(1)$_Y$ and fermions masses, and responsible for $m_W^2= m_Z^2\ctw^2$ at tree-level, 
  we have the relations \cite{preparation} 
  \footnote{The terms proportional to $Y_f$ arise from the  operator $ \partial_\mu h Z_\nu g'B^{\mu\nu}$    that, after field redefinitions,    can be rewritten as interactions in \eq{prim} and \eq{rest}. One has to have this in mind when estimating the size of the  coefficients.} 
\bea
\kappa_{WW}&=& \ctw^2\kappa_{ZZ}+s_{2\theta_W}\kappa_{Z\gamma}+\stw^2\kappa_{\gamma\gamma}\, ,\label{custoini}\\
{\ctw} g^h_{Z\!f\!f}&=& {\sqrt{2}}\, T_{3 f}\,  g^h_{W\!f\!f'}V_{\rm CKM}^\dagger-Y_f\, \delta g^h_{ZZ}/m_W
\ \ \ \ \text{for $f$= up-type fermion}\, ,\nonumber\\
{\ctw}g^h_{Z\!f'\!f'}&=& {\sqrt{2}}\, T_{3 f'}\, V_{\rm CKM}^\dagger g^h_{W\!f\!f'}-Y_{f'}\, \delta g^h_{ZZ}/m_W
\ \ \ \ \text{for $f'$= down-type fermion}\, ,
\label{custo}
\eea
where $T_{3 f}$ and  $Y_f$  are  respectively the 3-component isospin   and hypercharge of the fermion $f$, with   
$Q_f=T_{3 f}+Y_f$ the electric charge, and $V_{\rm CKM}$   the CKM quark-mixing matrix \cite{Peskin:1995ev}.
\eq{custoini} was  first derived in \cite{Contino:2013kra}.
A left-right parity $P_{\rm LR}$     \cite{Elias-Miro:2013mua}
can further restrict the coefficients \cite{preparation}:
\be
\kappa_{Z\gamma}=\frac{c_{2\theta_W}}{s_{2\theta_W}}\kappa_{\gamma \gamma}\, .
\ee
Similar expressions  are derived for the CP-violating counterparts.

We can also have a reduction of Higgs couplings due to  dynamical reasons.
For example,  in BSMs with a strongly-interacting Higgs,  we can neglect $\kappa_{ZZ,WW}$ in comparison with  $g^h_{VV}$ and $\delta g^h_{ZZ}$,  as the formers are associated to interactions that contain more derivatives and therefore are expected to be smaller in our $D_\mu/\Lambda$ expansion (see later a power counting for these couplings). Also 
in  "universal"  BSMs (as those in which  the  BSM states only couple to SM bosons and not to fermions) 
 we only have  three   relevant contact-interactions $ hV_\mu J^\mu$:
\be
g^h_{Z\!J_3}\, \frac{h}{v}Z_\mu J^\mu_{3}\ ,\ \ \ 
g^h_{Z\!J_Y}\,\frac{h}{v}Z_\mu J^\mu_Y\  ,\ \ \
g^h_{W\!J}\,\frac{h}{v}\left(W^+_\mu J^\mu_{W}+h.c.\right)\, ,
\ee
where $J^\mu_{3}$, $J^\mu_{Y}$ and  $J^\mu_{W}$ are  respectively the 3-component isospin,  hypercharge and charged  SM currents \cite{Peskin:1995ev}. Demanding also custodial invariance, we obtain
\be
 g^h_{Z\!J_3}=\frac{g^h_{W\!J}}{\ctw} \ ,\ \ \ \ g^h_{Z\!J_Y}=-\frac{\delta g^h_{ZZ}}{\ctw m_W}\, ,
\ee
that is equivalent to
\be
g^h_{W\!f\!f'}=g^h_{WJ}V_{\rm CKM}\ , \ \ \ \ \ 
g^h_{Z\!f\!f}= T_{3 f} \frac{{\sqrt{2}} g^h_{WJ}}{{\ctw}} -Y_{f}\frac{\delta g^h_{ZZ}}{\ctw m_W}\, .
\label{custo2}
\ee
\eq{custo2}, together with   \eq{custoini}, show that  universality and custodial symmetry reduce  \eq{rest}  to only 3 independent Higgs couplings, that we can take to be $\delta g^h_{ZZ}$, $\kappa_{ZZ}$ and $g^h_{WJ}$. 
This is  in accordance with \cite{Contino:2013kra}.

\section{The SM predictions for Higgs couplings}
In the SM the Higgs sector is  given by 
\be
{\cal L}^{\rm SM}_h=|D_\mu H|^2- (y_u  \bar Q_L \widetilde H u_R+y_d  \bar Q_L Hd_R+y_e \bar L_L He_R+h.c.)+\mu^2|H|^2-\lambda|H|^4\, ,
\ee
where the complex Higgs field $H$ is a $\bf 2_{1/2}$ of SU(2)$_L\times$U(1)$_Y$, $\widetilde H= i\sigma^2 H^*$,  and
\be
Q_L=\left(\begin{array}{c}u_L \\d_L\end{array}\right)\ ,\ \ \ \ \ \\ 
L_L=\left(\begin{array}{c}\nu_L \\ e_L\end{array}\right)\, .
\ee
When the Higgs gets a vacuum expectation value (VEV), 
$\langle H\rangle=(0\ v/\sqrt{2})^T$,  where $v\simeq 246$ GeV,
 the gauge bosons $W/Z$ and fermions  get a mass proportional to their coupling to the Higgs field.
Out of the 4 degrees of freedom in $H$, 3 corresponds to the would-be Nambu-Golstone bosons that 
become the longitudinal component of the $W$ and $Z$, and the 4th is the Higgs particle $h$.
In the SM all couplings of the Higgs are predicted as a function of particle masses.
We have, at tree-level, that the only nonzero couplings are
\be
g^h_{ff}=-\frac{gm_f}{2m_W}  \ , \ \ \  g^h_{VV}={gm_W}  \ , \ \ \  g_{3h}=-\frac{3g m^2_h}{2m_W}\, ,
\label{SMvalues}
\ee
that lead to the straight line of  Fig.~\ref{Higgscouplings}.
 The rest of the Higgs couplings  arise at the loop level; $\kappa_{GG}$ is mainly induced by the top loop, while
    $\kappa_{\gamma\gamma}$ and    $\kappa_{Z\gamma}$ are  generated by $W$ and top  loops,
as can be found for example in \cite{Gunion:1989we}.

\section{Higgs couplings  in an Effective Field Theory  approach to the SM}

Let us consider  BSMs  characterized by a mass-scale  $\Lambda$ much larger than the electroweak scale $m_W$, such that, after integrating out the BSM sector, we can make an expansion  not only in derivatives $D_\mu$ over $\Lambda$, as we did in previous sections, but also an expansion of SM fields over $\Lambda$.
 In this way we can obtain an Effective  Field Theory (EFT)   made of   local operators:
\footnote{This EFT also contains operators of dimension five, ${\cal L}_5$, but these induce neutrino masses
and therefore their coefficients must be very small (or their suppression scale $\Lambda$ very large). For this reason we neglect them here since they cannot play any role for Higgs physics at the TeV.}
\begin{equation}
{\cal L}_{\rm EFT}=\frac{\Lambda^4}{g^2_*}
{\cal L}\left(\frac{D_\mu}{\Lambda}\ ,\  \frac{g_* H}{\Lambda}\ ,\  \frac{g_{*}  f_{L,R}}{\Lambda^{3/2}}\ ,\ \frac{gF_{\mu\nu}}{\Lambda^2}\right)\simeq {\cal L}_4+{\cal L}_6+\cdots\, .
  \label{EFT}
\end{equation}
Here ${\cal L}_d$ denotes  the  term in the expansion made of local operators of dimension $d$,  while
  $g_*$  denotes a generic coupling, and  $g$ and $F_{\mu\nu}$ represent respectively the SM gauge couplings and field-strengths.   The Lagrangian in~\eq{EFT} is based on dimensional analysis and the 
dependence on the coupling $g_*$  is easily obtained when the Planck constant $\hbar$  is put back in place.
Indeed, working with units $\hbar\not=1$,  the couplings have dimensions $[g_*]=[\hbar]^{-1/2}$, 
while $[H]=L^{-1}\cdot [\hbar]^{1/2}$ and  the Lagrangian mass-terms $[\Lambda]=L^{-1}$.
This dictates the dimensionless expansion-parameters to be  $g_* H/\Lambda$ and $D_\mu/\Lambda$,
and that terms  in the Lagrangian that contains $n$ fields must  carry ${n-2}$ couplings to have the right dimensions.
This counting is  therefore  valid even if $g_*$ is not small. 
Although  we are using a generic coupling and mass-scale, $g_*$ and $\Lambda$,  it is clear
that this ought not to  be always the case. For example, for a strongly-interacting  light Higgs (SILH) \cite{Giudice:2007fh}
only the couplings of the Higgs  to the strong BSM sector are large  ($g_*\gg 1$ for the Higgs),
while SM  fermions  are assumed to have small couplings  ($g_*\sim \sqrt{y_f}$ for  fermions).

The Lagrangian terms   of ${\cal L}_4$ redefine  the  SM  (and have no physical impact),
while  ${\cal L}_6$ encodes the dominant BSM effects.
Therefore the study of the physical implications of ${\cal L}_6$ in  the physics of the SM is of great importance.
  There are different bases  used in the literature for  the set of independent $d=6$ operators in  ${\cal L}_6$.
  Although physics is independent of the choice of basis, it is clear that some bases
are better suited than others in order to extract the relevant information, {\em e.g.}, for Higgs physics.
  The first complete and non-redundant basis of dimension-6 operators  was given in  \cite{Grzadkowski:2010es}. 
  The virtue of that basis is that it is constructed with the maximum number  of operators made of fields 
  instead of using derivatives, following our approach for \eq{prim} and \eq{rest}.
 As we mentioned,   this can be useful when  estimating the size of the coefficients (see section below)  or 
looking for the dominant effects at  high-energies.
Nevertheless,  from a model-building point of view, it can  be more advantageous  to define  bases that
capture  in  few operators    the impact of the most interesting   BSM scenarios.
With this philosophy,  the SILH basis  was constructed in \cite{Giudice:2007fh},
and generalised to a complete ${\cal L}_6$  basis  in \cite{Contino:2013kra, Elias-Miro:2013mua}.
In this basis  "universal" BSMs   are encoded in few operators made only of SM bosons.
This has the virtue  of, for example,  having  a more direct connection between operator coefficients and 
the $S$ and $T$ parameters  \cite{Peskin:1990zt} that characterize the main electroweak effects of these BSMs.
This simplicity is not present in the basis of \cite{Grzadkowski:2010es} in which the equivalent of the $S$  and $T$ parameters
involve  vertex corrections \cite{Gupta:2014rxa} and then a less direct connection with the operator coefficients.
Another useful basis is given in \cite{Masso:2014xra}  with the   interesting  property of having a  one-to-one correspondence 
between   operators and  the most relevant physical  interactions measured at experiments.

In all the above mentioned  bases it is possible to separate  
the operators   into the following two groups: those that
could (in principle)  be induced {\it at tree-level} from integrating out
 heavy states with spin $\leq 1$  in renormalizable weakly-interacting BSMs,
and those operators that can only be induced {\it at the one-loop level} from these BSMs  \cite{Elias-Miro:2013mua, Einhorn:2013kja}.
This property is, however, not respected for bases  constructed  with the operators of 
\cite{Hagiwara:1993ck} where {\it tree} and {\it loop} operators are mixed.

The coefficients of ${\cal L}_6$,   referred as Wilson coefficients,  
are generated at the scale $\Lambda$ where 
they are generated  after integrating out the BSM heavy states.
The renormalization group evolution (RGE) from $\Lambda$  down to the electroweak scale, where they are supposed to be measured,  can give important corrections to the Wilson coefficients and mix them
\cite{Elias-Miro:2013gya,Elias-Miro:2013mua,Jenkins:2013zja}.
 For example, in supersymmetric theories
or composite Higgs models, where the Wilson coefficients can be determined (see below),
the RGE give us  the leading-log corrections to the predictions for the Higgs couplings 
at low-energy that can be significant  in certain cases \cite{Elias-Miro:2013mua}.

The full set of physical implications of ${\cal L}_6$ was  given
in \cite{Gupta:2014rxa}, where it was shown 
that  not all type of  interactions can be obtained  from ${\cal L}_6$ and, 
of the possible ones, not all of them are independent.
The set of independent couplings 
that are, at present, the experimentally best tested ones,
  were called  {\it primary} couplings.
The ones of the Higgs   are presented below.

\subsection{Primary Higgs couplings}

Among all dimension-6 operators  present in ${\cal L}_6$,  there are   few of them 
that  contribute {\it only} to  Higgs couplings  and  not to other   couplings (such as $Vff$) \cite{Elias-Miro:2013mua}.
These are the  set of independent dimension-6 operators
  constructed with $|H|^2$.
  The  CP-conserving  ones are
\footnote{Notice that the  operator    $|H|^2f\Dslash f$ can always be eliminated from the Lagrangian by  field redefinitions.}
\bea
&  |H|^2 \bar Q_L \widetilde Hu_R+h.c.\ ,\  \
  |H|^2 \bar Q_L Hd_R+h.c.\ ,\ \
  |H|^2 \bar L_L He_R+h.c.\ ,\ \
  & \nonumber\\
   &|H|^2 |D_\mu H|^2\ ,\ |H|^6\ , \ \ 
   |H|^2 G^{A\, \mu\nu}G^A_{\mu\nu}\ ,\ \ 
  |H|^2 B^{\mu\nu}B_{\mu\nu}\ ,\  \
  |H|^2 W^{a\, \mu\nu}W^a_{\mu\nu}\, ,&
\label{dim6primaries}
  \eea
where $W^a_\mu,B_\mu$ are the SU(2)$_L\times$U(1)$_Y$ gauge bosons.
 To see that, indeed,
 the above operators can {\it only} be probed by measuring  Higgs couplings,
 we  just have to  put the   Higgs field in the EWSB vacuum,  $|H|^2\to v^2/2$, 
 and realize that  the  resulting terms    are operators already present in the SM, {\it i.e.},
their  only  effect  is  a redefinition of the SM parameters.

The set of Higgs couplings that can be independently generated from \eq{dim6primaries}
are the  primary Higgs couplings \cite{Gupta:2014rxa}.  Their measurements  provide new probes to new physics only accessible  by Higgs physics.
  The number of  primary Higgs  couplings   must obviously coincide with the number of   Wilson coefficients associated with the operators of \eq{dim6primaries} (for the CP-conserving case).
  We have chosen  as primary Higgs  couplings those in \eq{prim},
as all of them can be independently generated from the operators of \eq{dim6primaries}.
We must be aware  however  that the correspondence is not one-to-one \cite{Elias-Miro:2013mua, Pomarol:2013zra}.
There is a certain freedom to choose the  set of  primary Higgs  couplings.
For example,  instead of  $\kappa_{\gamma\gamma}$ and $\kappa_{Z\gamma}$,
we could have taken $\kappa_{ZZ,WW}$, as these latter  can also receive independent contributions from \eq{dim6primaries}. 
The reason to choose \eq{prim} as  primary Higgs  couplings it is just  experimental:
they are the set of primary Higgs couplings  best measured at the LHC.

Similarly,  the CP-violating dimension-6 operators constructed with $|H|^2$ are
\bea
& i |H|^2 \bar Q_L \widetilde Hu_R+h.c.\ ,\  \
 i |H|^2 \bar Q_L Hd_R+h.c.\ ,\ \
 i |H|^2 \bar L_L He_R+h.c.\ ,&\nonumber\\
& |H|^2 G^{A\, \mu\nu}\widetilde G^A_{\mu\nu}\ ,\  \
  |H|^2 B^{\mu\nu}\widetilde B_{\mu\nu}\ ,\  \
  |H|^2 W^{a\, \mu\nu}\widetilde W^a_{\mu\nu}\, ,&
  \eea
that can independently generate  the set of  primary Higgs couplings of \eq{primCPV}.
Again, all these operators  for  $|H|^2\to v^2/2$ generate SM terms (that redefine SM  parameters)
and therefore  their physical  effects can only be seen in  Higgs physics.

The primary Higgs   couplings can enter at the quantum level in other non-Higgs observables.
For example, the CP-violating Higgs couplings can contribute at the loop-level to the neutron and  electron electric
dipole moment (EDM). The fact that we have excellent bounds on these EDMs, place indirect bounds on
these Higgs couplings. We must be aware however that  these bounds  are model-dependent, as 
there can be, in principle,  other BSM effects entering in the EDMs.

\subsection{Beyond the primaries}

The rest of CP-conserving Higgs couplings, beyond the primaries, are those of \eq{rest}  at the order we mentioned before.
 They can in principle be generated from operators in ${\cal L}_6$. 
 \footnote{At $O(hFff)$ we also have dipole-type interactions that can arise from ${\cal L}_6$. Their Wilson coefficients are however expected to be suppressed by SM  Yukawa-couplings 
(otherwise could largely contribute at the loop level to the SM fermion masses). These couplings are related to
fermion EDMs as can be found in \cite{Gupta:2014rxa}.}
 Nevertheless, it can be proven \cite{Elias-Miro:2013mua, Pomarol:2013zra}
that contributions from ${\cal L}_6$ to \eq{rest} 
are not independent from  contributions to  primary Higgs  couplings and   other electroweak  couplings.
Therefore they  can, in principle,  be  constrained by other experimental measurements.
As an example, consider the  operator $H^\dagger D_\mu H \bar e_R\gamma^\mu e_R$. This gives  a 
contribution to the Higgs coupling $g^h_{Z\!f\!f}$, but  it also contributes to the coupling $Z\bar e_Re_R$ 
that has been very-well measured at LEP, putting  strong bounds on possible BSM effects.   

 The explicit relations between the ${\cal L}_6$-contributions to  \eq{rest} and to other couplings were  explicitly  calculated in
  \cite{Pomarol:2013zra,Gupta:2014rxa,Masso:2014xra}
 assuming  family universality. Here we give these relations for the general case (derived at the tree-level) \cite{preparation}:
\begin{align}
\delta g^h_{ZZ}&=2g \ttw^2 m_W   \left(\ctw^2 {\delta g_1^Z}  -\delta \kappa_\gamma\right)\, , & \nn\\
g^h_{Z\!f\!f}&=2 {\delta g^Z_{f\!f}}-2 {\delta g_1^Z} (g^Z_{f\!f} c_{2 \theta_W}+g^\gamma_{f\!f} s_{2 \theta_W})+2  \delta \kappa_\gamma  Y_f 
\frac{e\stw}{\ctw^3} \, ,&
 g^h_{W\! f\!f'}  &=  2 {\delta g^W_{f\!f'}}-2 {\delta g_1^Z} g^W_{f\!f'} c^2_{\theta_W}\, ,
\label{predictions} \\
\kappa_{ZZ}&=\frac{1}{\ctw^2}  {\delta \kappa_\gamma}+2\frac{c_{2\theta_W}}{s_{2\theta_W}} \kappa_{Z \gamma} + { \kappa_{\gamma \gamma}}   \, , &
\kappa_{WW}& ={\delta \kappa_\gamma}+{ \kappa_{ Z \gamma }}+{\kappa_{\gamma \gamma }}  \, ,
\label{predictions1}
\end{align}
with
\be
\delta g^W_{f\!f'}= \frac{\ctw}{\sqrt{2}}\left(\delta g^Z_{f\!f}V_{\rm CKM}-V_{\rm CKM}\delta g^Z_{f'\!f'}\right)\ \ \text{for}\  \ 
f=f_L
\, ,
\ee
and where
\be
g^\gamma_{f\!f}=e Q_f\ , \ \ \ 
g^Z_{f\!f}=\frac{g}{\ctw}\left( T_{3 f}-Q_f \stw^2 \right)\ ,\ \  \ \ g^W_{f\!f'}=\frac{g}{\sqrt{2}}V_{\rm CKM},\, 0\ \ \text{resp. for}\  \ f=f_L,f_R\ ,
\label{SMc}
\ee
are the $\gamma$, $Z$ and $W$ couplings to fermions in the SM. Flavor  indices are again implicit.
We have also defined by   $\delta g^Z_{f\!f}$ ($\delta g^W_{f\!f'}$) the  BSM corrections to the $Z$ ($W$) couplings to fermions:
\be
\Delta {\cal L}^V_{ff} =\frac{\delta g^Z_{f\!f}}{2}\left(Z_\mu J^\mu_{N}+h.c.\right)+
\delta g^W_{f\!f'}\left(W^+_\mu J^\mu_{C}+h.c.\right)\, ,
\ee
while  $\delta g_1^Z$  is the correction to the  $ZWW$ coupling   and
 $\delta \kappa_\gamma$ parametrizes  BSM contributions to the  EDM of the $W$, following the notation of \cite{Hagiwara:1993ck} for anomalous triple gauge couplings (TGC):
\be
\Delta {\cal L}_{3V} = i g \ctw \delta g_1^Z \left[ 
 \, Z^{\mu}  \left( W^{+\, \nu}  W^-_{\mu\nu} 
  -h.c.\right)+ Z^{\mu\nu} W^+_\mu  W^-_\nu\right]
   + i e \delta \kappa_\gamma \left[ (A^{\mu\nu}-\ttw Z^{\mu\nu}) W^+_\mu  W^-_\nu  \right] \, .
\label{V3}
\ee
Following \cite{Gupta:2014rxa},
we have chosen to   work in the mass-eigenstate basis  within a  parametrization in which  kinetic terms and masses  do not receive corrections and then take the SM values. All BSM effects are in couplings.
We think this is the most convenient parametrization of BSM effects
due to the straightforward  connection between couplings and physical processes, that in most of the cases is a one-to-one
correspondence. The SM input parameters can be taken to be $\alpha_{\rm EM}$, $m_Z$ and $m_W$ that,
in our parametrization,  do not have BSM corrections, as opposed to $G_F$ that  receive corrections from 4-fermion interactions.
We remark again  that the predictions \eq{predictions}  and \eq{predictions1} are derived at the tree-level and
only apply to BSM effects coming from ${\cal L}_6$. There are also  SM contributions   to these couplings at the loop level, 
that can be as important as new-physics contributions,  and must be incorporated accordingly.

\eq{predictions}  and \eq{predictions1}  are important  results.
They  show that all Higgs couplings of  \eq{rest}  can be written  as a function  of BSM effects  to two primary Higgs  couplings 
($\kappa_{\gamma\gamma}$, $\kappa_{Z\gamma}$),
 $Z/W$ couplings to SM fermions ($\delta g^Z_{f\!f}$,  $\delta g^W_{f_R\!f'_R}$), and two TGC ($\delta g_1^Z$, $\delta \kappa_\gamma$). 
Experimental bounds on  $\kappa_{\gamma\gamma,Z\gamma}$ are already at the per-cent level \cite{Pomarol:2013zra},
while  $Z/W$ couplings have also been experimentally contrained, mostly from  LEP and SLC \cite{ALEPH:2005ab,LEP2wwzOLD}  (with  Tevatron providing an  accurate measurement of the $W$-mass).
One finds that bounds on  $\delta g^Z_{f\!f}$ are quite strong, at the per mille-level in most of the cases,
but bounds on $\delta g_1^Z$ and $\delta \kappa_\gamma$ are much weaker \cite{ewpt}.
Therefore, at present,
we can already  derive, using \eq{predictions}  and \eq{predictions1},
   relevant model-independent bounds on  the Higgs couplings of  \eq{rest} \cite{Pomarol:2013zra}.

In the case of custodial-invariant universal BSMs, \eq{predictions} reduces to 
\begin{align}
\delta g^h_{ZZ}&=2g \ttw^2 m_W   \left( \ctw^2{\delta g_1^Z}  -{\delta \kappa_\gamma}+
 \widehat S \right)\, , & \nn\\
g^h_{Z\!f\!f}&= -2T_{3 f}\,   \delta g_1^Z g c_{\theta_W}-Y_{f}\frac{\delta g^h_{ZZ}}{\ctw m_W}
 \ ,&\ \ \ 
 g^h_{W\! f\!f'}  &=-2 {\delta g_1^Z} g^W_{f\!f'} c^2_{\theta_W}\, ,\ \ \ 
\label{predictions2}
\end{align}
where $\widehat S$ is, up to a normalization constant  \cite{Barbieri:2004qk},   the 
$S$-parameter \cite{Peskin:1990zt}.  As expected, \eq{predictions2} and \eq{predictions} fulfill \eq{custoini} and \eq{custo2},
and  $g^h_{Z\!f\!f}$ is fully determined by the custodial symmetry as a function of  $ g^h_{W\! f\!f'}$ and $\delta g^h_{ZZ}$.

The  CP-violating non-primary Higgs couplings, \eq{restCPV}, are also not independent but related to other  couplings. We have
\begin{align}
\tilde g^h_{Z\!f\!f}&=
2 {\delta \tilde g^Z_{f\!f}}\, ,&
\tilde g^h_{W\! f\!f'}  &= 2  {\delta \tilde g^W_{f\!f'}}\, ,\nonumber \\
\widetilde\kappa_{ZZ}&=\frac{1}{\ctw^2}  {\delta \widetilde\kappa_\gamma}+2\frac{c_{2\theta_W}}{s_{2\theta_W}}\widetilde \kappa_{Z \gamma} + { \widetilde\kappa_{\gamma \gamma}}  \, , &
\widetilde\kappa_{WW}& ={\delta \widetilde\kappa_\gamma}+{ \widetilde\kappa_{ Z \gamma }}+{\widetilde\kappa_{\gamma \gamma }}\, ,
\label{predictionsCPV}
\end{align}
where 
\be
\delta \tilde g^W_{f\!f'}= \frac{\ctw}{\sqrt{2}}\left(\delta \tilde g^Z_{f\!f}V_{\rm CKM}-V_{\rm CKM}\delta\tilde  g^Z_{f'\!f'}\right)\ \text{for}\  \ f=f_L\, ,
\ee
with  $\delta \tilde g^Z_{f\!f}$  and $\delta \tilde g^W_{f\!f'}$ defined as
\be
\Delta \tilde {\cal L}^V_{ff} =\frac{\delta \tilde g^Z_{f\!f}}{2}\left(iZ_\mu J^\mu_{N}-h.c.\right)+
\delta \tilde g^W_{f\!f'}\left(iW^+_\mu J^\mu_{C}-h.c.\right)\, ,
\ee
and  $\widetilde\kappa_\gamma$ being  the CP-violating TGC:
\be
\Delta {\cal L}_{3\widetilde V} = 
    i e \delta \widetilde\kappa_\gamma \left[ (\widetilde A^{\mu\nu}-\ttw \widetilde Z^{\mu\nu}) W^+_\mu  W^-_\nu  \right] \, .
\label{V3CPV}
\ee
The predictions \eq{predictions}, \eq{predictions1} and \eq{predictionsCPV}  rely on  the  (quite plausible) hypothesis
that  the leading SM deviations arise from ${\cal L}_6$.
Finding experimental evidence for deviations from these predictions,
   would mean that nature does not fulfil  this hypothesis:
   either because there are  light BSM states ($\Lambda\lesssim m_h$), 
the composite-scale of the Higgs is  low  ($\Lambda\sim g_* v$), that is  equivalent to  say that $h$ cannot be identified within the SM doublet, or that there are other sources of EWSB independent of $\langle H\rangle$ \cite{SSSB}.

\subsection{Power counting for  Higgs couplings}
\label{pc}
It can be   useful to estimate the size of the contributions to the effective Higgs couplings arising  from generic  BSMs.
As it is clear from the expansion in \eq{EFT},  the  coefficients in \eq{prim} and \eq{rest}
can have different dependence with $g_*$.  
The  Higgs couplings that can  receive   the largest power of $g_*$  are $g_{3h}$ and $g^h_{ff}$ where
\be
\delta g_{3h}\sim \frac{g^4_*v^3}{\Lambda^2}\ ,\
\delta g^h_{ff}\sim \frac{g^3_* v^2}{\Lambda^2}\, .
\label{largest}
\ee
For $g_*\gg 1$, \eq{largest}  can give $O(1)$ corrections to $g_{3h}$ and $g^h_{ff}$, 
even after demanding $g_*^2v^2/ \Lambda^2\ll 1$ necessary to 
make the expansion \eq{EFT} valid.
Nevertheless, in theories where the Higgs mass is protected by a symmetry, as it happens in theories that solve the hierarchy problem such as supersymmetry or composite Higgs models, the contributions to $g_{3h}$  are also expected to 
be protected and  then proportional to $m^2_h/v^2\sim \lambda$.
Also  it is natural to expect that  chirality protects terms proportional to $\bar f_Lf_R$, at least by a Yukawa coupling $y_f\sim m_f/v$, otherwise corrections to fermion masses would be too large.
For this reason, it is more natural to assume that the corrections to these Higgs couplings are of order
\be
\delta g_{3h}\sim \lambda v  \frac{g^2_*v^2}{\Lambda^2}\ ,\
\delta g^h_{ff}\sim y_f \frac{ g^2_* v^2}{\Lambda^2}\, ,
\label{nda1}
\ee
that  potentially give  relative corrections  of $O(g^2_*v^2/\Lambda^2)$.
 At the same order,  we also have
\be
\delta g^h_{VV}\sim g^2v\frac{g^2_* v^2}{\Lambda^2}\, ,
\label{nda2}
\ee
and 
\be
\delta g^Z_{f\!f},\delta g^Z_{1}\sim g\frac{g^2_*v^2}{\Lambda^2}\, .
\label{nda3}
\ee
Finally, couplings coming from a derivative (or field-strength) expansion, the $\kappa_i$, are expected to scale as
\be
\kappa_i  \sim \frac{g^2v^2}{\Lambda^2}\, .
\label{ndakappa}
\ee
Nevertheless, in renormalizable  BSMs these coefficients can only be induced at the loop-level 
  and therefore  expected to be 
  \be
\kappa_i  \sim \frac{g_{\rm *}^2}{16\pi^2}\frac{g^2v^2}{\Lambda^2}\, .
\ee
Indeed,  it can be shown   \cite{Giudice:2007fh,Elias-Miro:2013mua}
  that the $\kappa_i$ cannot be generated {\it at tree-level} 
 from integrating out scalars, fermions and vector bosons  in renormalizable theories.

The above estimates are useful  to determine which are the most sizeable BSM corrections to the Higgs couplings.
For example,  in theories in which the Higgs is strongly coupled, 
 the largest  corrections are those of \eq{nda1}  and \eq{nda2} that depend quadratically in the strong coupling $g_*\gg 1$\cite{Giudice:2007fh}. If also the  SM fermions are strongly-coupled,   \eq{nda3} can also give similar size corrections.  
It is also important to  remark  that even for theories in which the field expansion in \eq{EFT} is not  valid  ({\em e.g.}, when $g_*v\sim\Lambda$), the power counting for  Higgs couplings  given here is  expected to be correct.
In particular, the above estimates  are in accordance with the NDA analysis of \cite{Manohar:1983md} proposed for QCD.

  For the  non-primary Higgs couplings we have the estimates
\be
\frac{\delta g^h_{ZZ}}{g^2v}, \frac{g^h_{Z\!f\!f}}{g},\frac{g^h_{W\!f\!f'}}{g}\sim  \frac{g^2_*v^2}{\Lambda^2}\ \ \ \text{and}\ \ \  \ 
\kappa_{ZZ},\kappa_{WW}\sim \frac{g^2v^2}{\Lambda^2}\, ,
\label{estimatehV}
\ee
in agreement with the relations in \eq{predictions}.
Similar estimates follow for  CP-violating Higgs   couplings.

\section{Experimental  determination of the effective Higgs couplings}

\begin{figure}
\centering
\hskip-.6cm\includegraphics[width=.5\linewidth]{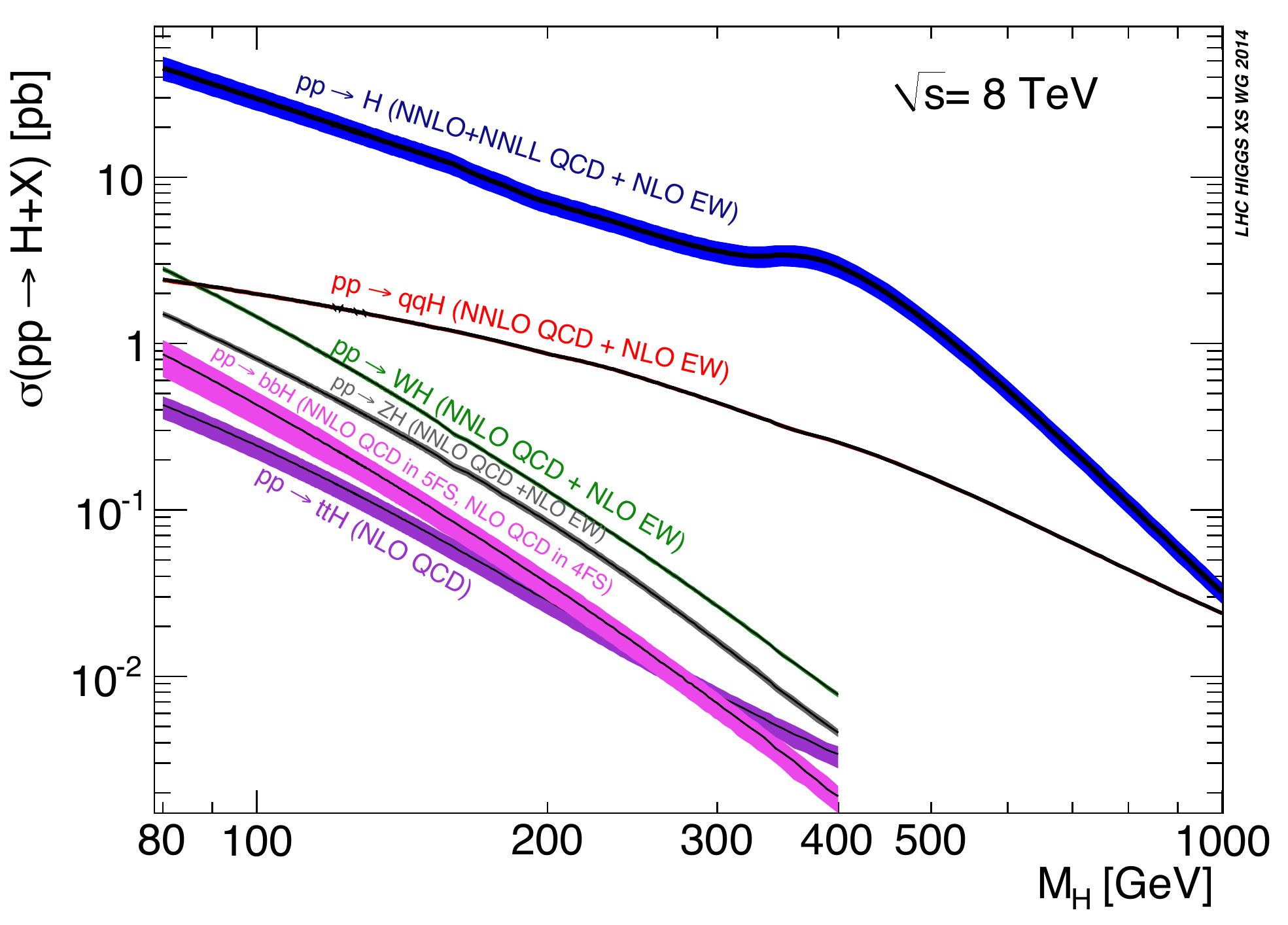}\hskip1.5cm
\includegraphics[width=.38\linewidth]{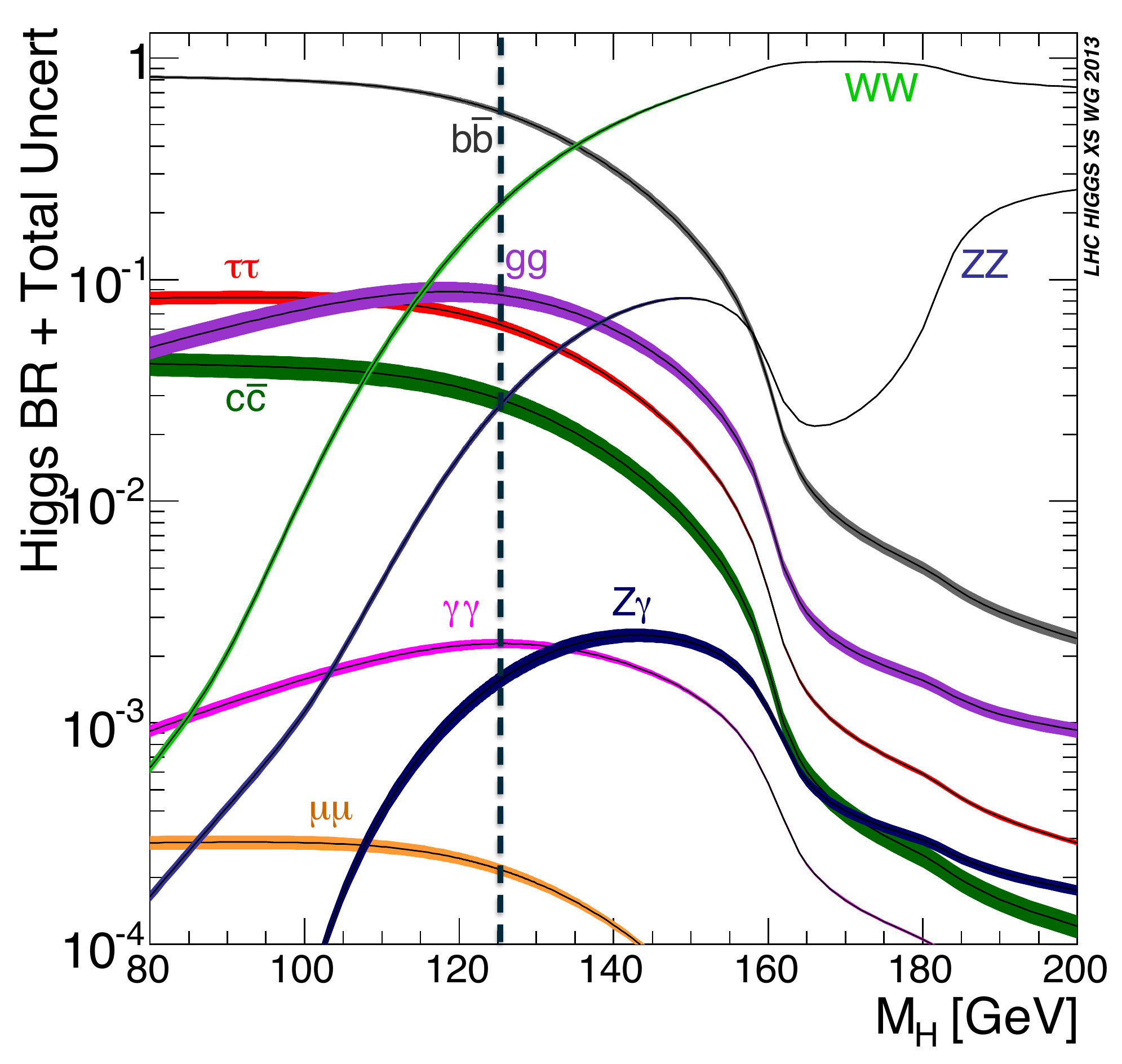}
\caption{Predictions for the main Higgs production cross-sections and Higgs BR in the SM \cite{LHCHXSWG}.}
\label{SMpredictions}
\end{figure}

\begin{figure}
\centering\includegraphics[width=.45\linewidth]{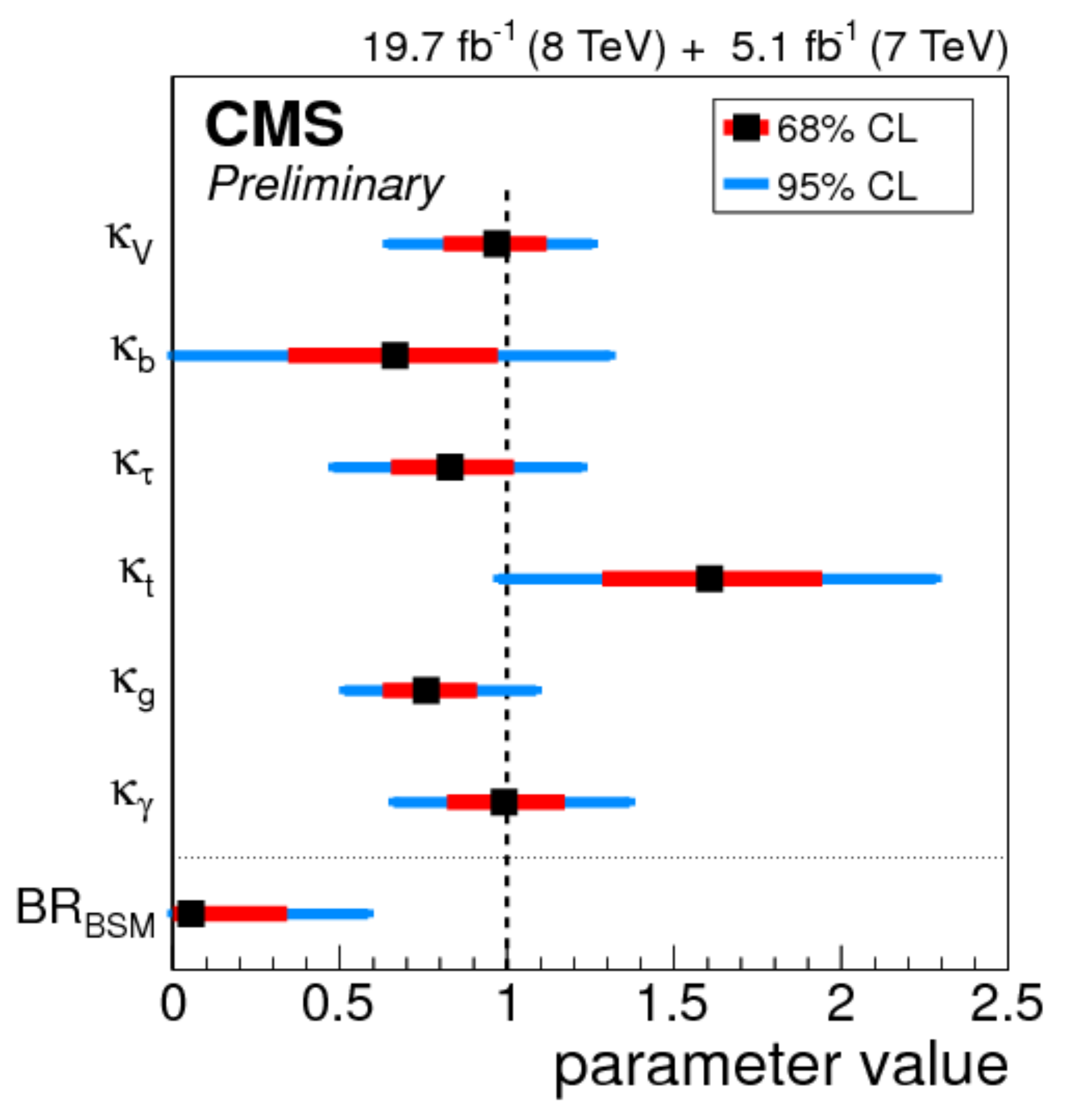}
\caption{Fit of  6 primary Higgs  couplings from CMS \cite{CMS:2014}. 
 Notation: $\kappa_V\equiv g^h_{VV}/g^{h\, \rm SM}_{VV}$,
$\kappa_f\equiv g^h_{ff}/g^{h\, \rm SM}_{ff}$,
$\kappa_g\equiv \kappa_{GG}/\kappa^{\rm SM}_{GG}$
and $\kappa_\gamma\equiv \kappa_{\gamma\gamma}/\kappa^{\rm SM}_{\gamma\gamma}$;
 loop effects in $\kappa_{GG}$ and $\kappa_{\gamma\gamma}$ are not included \cite{CMS:2014}.}
\label{CMS}
\end{figure}

The  primary  Higgs couplings  can be determined by searching for the Higgs  
through the  different  production mechanisms  and  decays.
The main Higgs production mechanisms at the LHC are
\bea
 \text{Gluon fusion:} &&GG\to h\, ,\nn\\
 \text{$Vh$-associated production:} &&q\bar  q\to Vh\, ,\nn\\
\text{Vector boson fusion (VBF):} &&  qq\to qqVV^*\to qq h\, ,\nn\\
\text{$htt$-associated production:}   && GG\to t\bar th\, ,
\label{rates}
\eea 
while the most important   Higgs branching ratios (BR) are 
\be
BR(h\to b\bar b)\ ,\
BR(h\to \tau\bar \tau)\ ,\
BR(h\to Vf\bar f )\ ,\
BR(h\to \gamma \gamma)\ ,\
BR(h\to Z\gamma)\, .
\label{brs}
\ee
The predictions for a  SM Higgs are given in Fig.~\ref{SMpredictions}.
 The Higgs mass can be mainly determined  from the Higgs decay to $\gamma\gamma$  and    $Zff$
 that allows to obtain
 \bea
 m_h&=&125.03\begin{array}{c}+0.26 \\-0.27\end{array}(stat.)
 \begin{array}{c}+0.13 \\-0.15\end{array} (syst.)\ \  \text{GeV\ \ \ \  from CMS}\, ,\nn\\  
 m_h& =& 125.36 \pm 0.37\ (stat.) \pm 0.18\ (syst.)\ \  \text{GeV\ \ \ \   from ATLAS}\, .  
\eea  
At the LHC one can combine the  different Higgs production mechanisms and  BR   of \eq{rates} and \eq{brs}
to determine   7 primary Higgs  couplings: $g^h_{ff}$ $(f=t,b,\tau)$,   $g^h_{VV}$, 
  $\kappa_{GG}$,  $\kappa_{\gamma\gamma}$ and  $\kappa_{Z\gamma}$. 
\footnote{We note   that  $g^h_{tt}$ and $g^h_{VV}$   also affect $BR(h\to \gamma \gamma/Z\gamma)$ and $\sigma(GG\to h)$ 
at the one-loop level \cite{Giudice:2007fh}.}
    The CMS   fit of six of the  primary  Higgs couplings is shown in Fig.~\ref{CMS}, where other Higgs couplings
have been set     to zero.
        \footnote{The ATLAS results are not shown here since the fit is performed only for    few primaries at each time instead of   a global fit to all of them\cite{ATLAS14}. For a combination of  ATLAS and CMS data see, for example, \cite{Falkowski:2013dza,Englert:2014uua}
.}
The fit shows a good agreement with the SM predictions and no sing of new-physics.
The implications of these  measurements  in particular BSMs will be discussed in the next section.
The primary coupling $\kappa_{Z\gamma}$ has  not been included in the fit of Fig.~\ref{CMS}, 
but one can use the experimental bound $BR(h\to Z\gamma)/BR(h\to Z\gamma)_{\rm SM}\lesssim 10$ \cite{Chatrchyan:2013vaa}  to derive the constraint $-0.01\lesssim \kappa_{Z\gamma}\lesssim 0.02$ \cite{Pomarol:2013zra}.
The fact that in the SM   $h\to Z\gamma$  arises at the one-loop level, and therefore has a small branching fraction
$BR(h\to Z\gamma)\sim 0.15\%$, makes this BR very sensitive  to new-physics;
it probably provides  the last  chance to find  large BSM  effects in  SM Higgs couplings.

Among the remaining  primary Higgs couplings to be measured
 we have $g_{3h}$. Its  determination however will be very   difficult since
 it requires  to search for double-Higgs production $pp\to hh$ that has small rates \cite{Contino:2010mh}. 
Also  Higgs couplings to light fermions $g_{ff}^h$ (beyond the 3rd family)
are going to be difficult to measure since  we expect these couplings to be proportional to $m_f/m_W$
(see \eq{SMvalues} and  \eq{nda1}), giving then very small BR.
For example, for the case of the muon, that is probably  the most accessible, we have in the SM
$BR(h\to\mu\mu)\sim 0.02\%$. Therefore a  high   luminosity at the LHC run  2 will be needed to measure this coupling.
 Flavour-violating Higgs couplings in  $g^h_{ff}$ can also  be accessible through Higgs decays.
This is particularly interesting for  theories of flavour in which  Yukawas are generated from the mixing of the   SM fermions with 
  heavy  BSM states. The strength  of these mixings are expected to be $\sim\sqrt{m_{f_i}/v}$, and therefore predicting
   $g^h_{f_i f_j}\sim \sqrt{m_{f_i} m_{f_j}}/v$ that can lead to   sizeable flavour-violating Higgs decays. 
In particular,  one has  $BR(h\to \tau \mu) \sim m_{\mu}/ m_{\tau}\times BR(h\to \tau \tau)\sim 0.4\%$
that    is quite close to the present experimental bound $BR(h\to \tau \mu)<1.57\%$  \cite{CMS:mutau}.
 Finally, most of the CP-violating Higgs couplings are poorly  measured since they 
appear  quadratically   in  production rates and BR since the   interference terms with the SM contributions vanish.
\footnote{Since in the SM the
$hGG$, $h\gamma\gamma$ and $hZ\gamma$ couplings are small (as they arise at the one-loop level),
the interference terms  are   also small, and the corresponding bounds on   CP-conserving and CP-violating couplings, $\kappa_i$ and $\widetilde\kappa_i$,  are comparable.}
Kinematical differential distributions can be used to  measure  these couplings \cite{CMS:xwa}, and alternative
methods have been recently proposed in \cite{Delaunay:2013npa}.
Nevertheless, indirect bounds on most of these couplings are very strong (see for example \cite{McKeen:2012av}
for bounds on $\widetilde\kappa_{\gamma\gamma}$    from  EDMs), making 
 difficult  to believe that Higgs CP-violating  couplings  are sizeable.
The exception is  probably $\delta\tilde  g^h_{\tau\tau}$ whose  bounds are not so strong
and could have possible impact in CP-violating Higgs decays.

\begin{figure}
\centering\includegraphics[width=.4\linewidth]{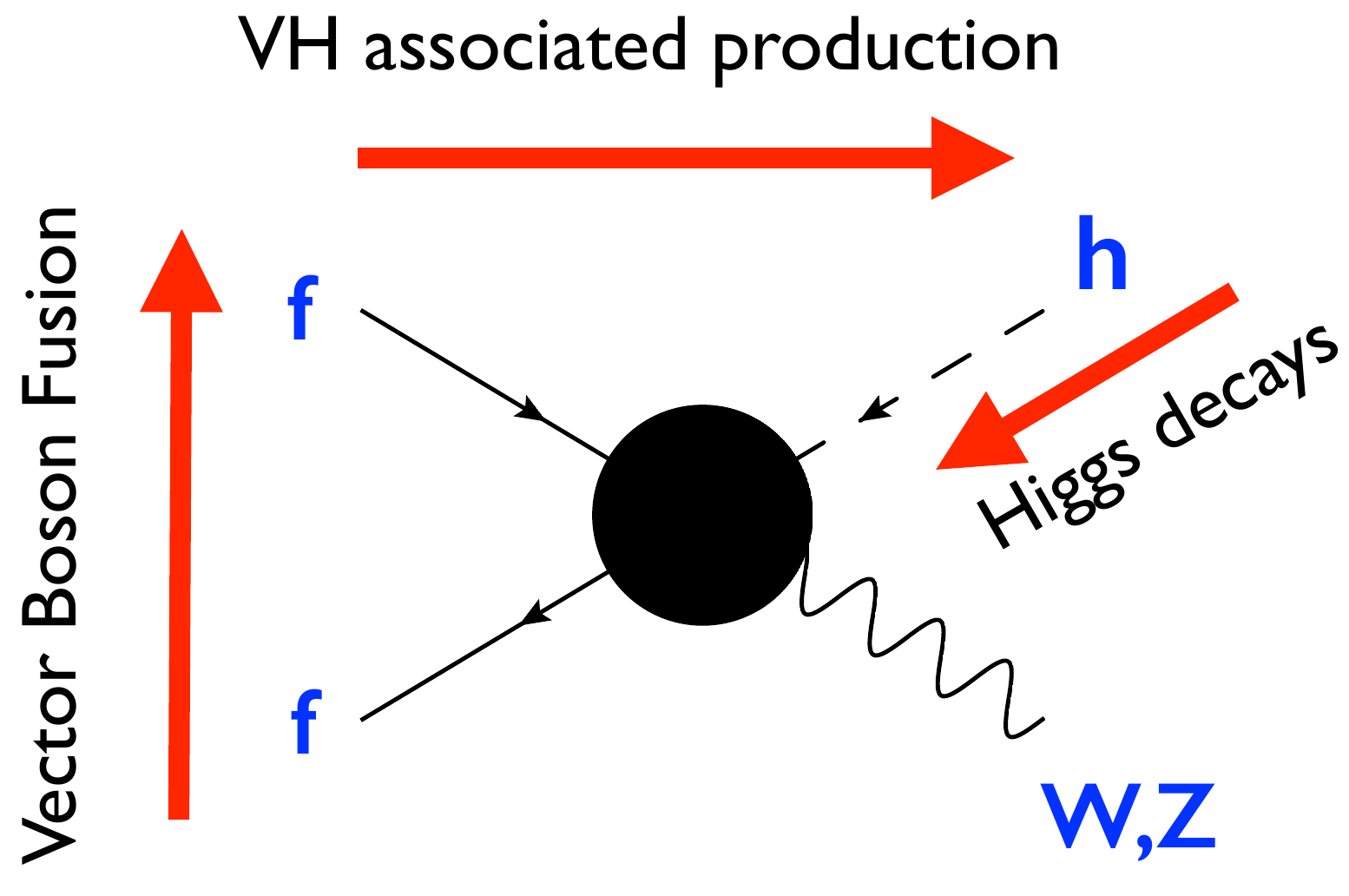}
\caption{The form-factor $hVff$, that as a function of the effective Higgs couplings is given in \eq{hVff},
can be tested in three different Higgs processes at the LHC:
either in Higgs decays $h\to Vff$,  in $Vh$-associated production or in the VBF-like process $pp\to qqV/qqVV^*\to qqh$.}
\label{hVffprocess}
\end{figure}

The experimental full extraction  of all   Higgs couplings, including the non-primary ones, \eq{rest} and \eq{restCPV}, is
a difficult task.
The best way to disentangle the effects of $\delta g^h_{ZZ}$,
$\kappa_{ZZ,WW}$ and   $g^h_{V\!f\!f}$ ($V=Z,W)$, as well as their CP-violating counterparts,
is by looking for  modifications in   differential distributions  in Higgs  processes.
The most relevant ones are 
the  Higgs decays $h\to Vff$,  the  $Vh$-associated production and  the VBF-like process  $pp\to qqV/qqVV^*\to qqh$.
All of them arise from the   $hVff$ amplitude (see Fig.~\ref{hVffprocess})  given by (neglecting fermion masses)
\be
 \mathcal{M}_{hV\!f\!f}(q,p) =\frac{1}{v} \epsilon^{*\mu}(q) \,  J_V^\nu(p) \left[ \, A^V  \, \eta_{\mu\nu} \, + \, B^V \,\left( p\cdot q\, \eta_{\mu\nu} -p_{\mu}\, q_{\nu}\right)  +C^V\epsilon_{\mu\nu\rho\sigma}p^\rho q^\sigma\right] \, ,
\label{hVff}
\ee
where $q$  and $p$ are respectively  the total 4-momentum of  $V$ and 
the fermion pair in $J_V^\mu=J_{N}^\mu, J_{C}^\mu$ for $V=Z,W$,
and $\epsilon^\mu$ is the polarization 4-vector of $V$.
We have defined
\be
A^V  = a^V+\widehat a^V\frac{  m_V^2}{p^2-m_V^2}\ ,\ \ \ \ \ \ 
 B^V =  b^V\frac{1}{p^2-m_V^2} +\widehat b^V\frac{1}{p^2}\ ,\ \ \ \ \ \ 
 C^V =  c^V\frac{1}{p^2-m_V^2} +\widehat c^V\frac{1}{p^2}\, ,
\label{ABC}
\ee
with $\widehat b^W,\widehat c^W=0$, and where
\begin{align}
a^Z&=\delta  g^h_{Z\!f\!f}+i\delta \tilde g^h_{Z\!f\!f}\, ,&
 a^{W} &=\delta  g^h_{W\!f\!f'}+i\delta \tilde g^h_{W\!f\!f'}\, ,\nonumber \\
\widehat a^Z&=2 g^Z_{f\!f}\left(1+\frac{\delta g^h_{VV}+\delta g^h_{ZZ}}{gm_W} \right)\, , & 
\widehat a^{W}& =2  g^W_{f\!f'}\left(1+\frac{\delta g^h_{VV}   }{gm_W}\right)\, ,  \nonumber\\
b^Z&=-2  g^Z_{f\!f}  \kappa_{ZZ} \, , &
b^{W}& =-2 g^W_{f\!f'}   \kappa_{WW}  \, ,\nonumber \\
\widehat b^Z&=-2 eQ_f \kappa_{Z \gamma}\, ,   &        \nonumber\\ 
c^Z&=-2  g^Z_{f\!f}  \widetilde\kappa_{ZZ} \, , &
c^{W}& =-2 g^W_{f\!f'}   \widetilde\kappa_{WW}  \, ,\nonumber \\
\widehat c^Z&=-2 eQ_f \widetilde\kappa_{Z \gamma}\, .   &         
\label{asbs}
\end{align}
From the differential distributions of the decay products in $h\to Vff$, one can put bounds 
on the coefficients of \eq{ABC} and, consequently, on  non-primary Higgs couplings. Nevertheless,  
we still have poor statistics  and  bounds on  Higgs couplings are almost irrelevant
 unless we turn on one by one  \cite{CMS:xwa}.
At present, the most promising way to obtain significant bounds  in some of the Higgs couplings of \eq{rest} is,
as we will discuss below, by  measuring them  at the LHC high-energy regime, for example in the $Vh$-associated Higgs production where the effects of some of these couplings are enhanced. 

Since  primary Higgs couplings  predict equal deviations
in the $hZff$ and $hWff$ physical amplitudes (normalized to their SM values), 
measuring  a relative deviation between these two  would provide evidence for
 non-primary Higgs couplings.
At the LHC  this relative deviation is parametrized by  $\lambda_{WZ}-1$ \cite{CMS:2014,ATLAS14}
that at  present  does not show  any evidence  of being different from zero; from the experimental
data we have   $-0.35<\lambda_{WZ}-1<0.08$ \cite{ATLAS14}.
This quantity is predicted in the SM EFT of \eq{EFT} to be
 \cite{Pomarol:2013zra}
 \be
\lambda_{W Z}^2-1
\simeq 0.6 \delta g^Z_1 - 0.5  \delta \kappa_\gamma -0.7 \kappa_{Z\gamma}\, ,
\ee
where we have used Eqs.~(\ref{predictions})-(\ref{predictions1}), neglecting $\kappa_{\gamma\gamma}$ and $\delta g^{Z,W}_{f\!f}$, since they are experimentally  constrained to be less than  $10^{-2}-10^{-3}$.

\subsection{Towards the high-energy regime}

One of the most interesting perspectives  at the LHC run 2  is the access to  physical processes at much higher energies.
This can be used to probe  Higgs production mechanism or  off-sell Higgs mediated processes
in a regime in which the  effects of some anomalous Higgs couplings can be  enhanced by  factors $E^2/\Lambda^2$.
As an example, let us consider the associated Higgs production, $pp\to Vh$.
As it is clear from  \eq{ABC}, at high-energies, $E\gg m_V$,  the coefficient $a^V$ dominates the amplitude.
Thanks to our  parametrization  for Higgs couplings, this coefficient is in one-to-one correspondence with 
the contact-interaction $g^h_{V\!f\!f}$.
Indeed,  at the partonic level, we have
\be
\sigma (qq\to hV)\Big|_{\hat s\gg m^2_h} =\ \sigma(qq\to hV_L)_{\rm SM}\left(1+\frac{g^h_{V\!f\!f}}{g_{f\!f}^V} \frac{\hat s}{m^2_V}+\dots\right)\, .
\label{largeE}
\ee
By looking at   high  invariant-masses for $hV$, it is possible 
to put important  bounds on $g^h_{V\!f\!f}$ \cite{Ellis:2014dva,Biekoetter:2014jwa}.
Nevertheless, one has to be careful  that one is  not probing these couplings at
energies above $\Lambda$ where an expansion in $\hat s/\Lambda^2$ would  not be valid.
To address  this issue, the power-counting of section~\ref{pc}  is crucial.
Using \eq{estimatehV}, we can write $g^h_{V\!f\!f}\equiv g c^h_{V\!f\!f} g^2_* v^2/\Lambda^2$
where  $c^h_{V\!f\!f}$ is a coefficient $O(1)$.
Now, experimentally,
due to the lack of  experimental accuracy in the measurement of   $pp\to Vh$ at the LHC, we can only bound  at present  high-energy   deviations from the SM  to be  less than $O(1)$ \cite{Ellis:2014dva,Biekoetter:2014jwa}, that is equivalent to say, using \eq{largeE},
\be
\frac{g^h_{V\!f\!f}}{g_{f\!f}^V} \frac{\hat s}{m^2_V}< O(1) \ \ \to \ \ 
c^h_{V\!f\!f}\lesssim \frac{\Lambda^2}{\hat s}\frac{g^2}{g^2_*}\,  .
\label{bound}
\ee
To guarantee the validity of the  expansion in ${\cal L}_h$, we must stay in the regime ${\Lambda^2}/{\hat s}\gg 1$.
Therefore the experimental bound \eq{bound} can only be restrictive (and  useful)  for strongly-interacting BSMs in which $g_*\gg g$.  In these  scenarios we can safely use the $hV$-production high-energy   data
to   obtain bounds on $g^h_{V\!f\!f}$ at the per-cent level \cite{Biekoetter:2014jwa}.
In models in which, in addition, the expansion  of \eq{EFT} is valid, 
 bounds on $g^h_{V\!f\!f}$
can be translated into bounds on $\delta g_1^Z$. 
Indeed, 
we  have from  \eq{predictions}, after
 neglecting $\delta g^Z_{f\!f}$ due to the strong constraints from LEP, and 
 neglecting $\delta \kappa_\gamma$ since this does not grow with $g_*^2$ \cite{Gupta:2014rxa},
\be
{\delta g_1^Z} \simeq  - \frac{g c^h_{Z\!f\!f}}{2(g^Z_{f\!f} c_{2 \theta_W}+e Q_f s_{2 \theta_W})}\frac{g^2_*v^2}{\Lambda^2} \simeq  - \frac{g c^h_{W\! f\!f'}}{ 2  g^W_{f\!f'} c^2_{\theta_W}}\frac{g^2_*v^2}{\Lambda^2}\, .
\ee
From the experimental data at the high-energy regime of the $hV$-associated production
we obtain   \cite{Biekoetter:2014jwa}
 \be
 -0.01 <\delta g^Z_1< 0.04\ \ \ \ \ (95\%\ \text{CL})\, .
 \ee
This is as competitive as the one
obtained from  anomalous TGC  at LEP \cite {LEP2wwzOLD} and at the   LHC \cite{ATLAS:2012mec}.

\subsection{Invisible Higgs decay}

We  have assumed so far that there are no more light particles than those of the SM.
If there  were new light states to which the Higgs could decay to, all the Higgs BRs
would be reduced, changing the fit of the Higgs couplings \cite{Chang:2008cw}.
There are well-motivated BSMs where the Higgs can decay  invisibly.
An example is given in \cite{Riva:2012hz}  where the Higgs can decay to a gravitino and neutrino
that interact  so weakly that escape from detection.
 Also in certain models the Higgs can decay to   dark matter that, being  stable and EM neutral, also escape from detection.
\footnote{Alternative effects from new  light physics can be found in \cite{light}.}
There are direct searches for Higgs decaying  invisibly based on 
looking for missing energy plus a $Z/W/\gamma/jet$. The CMS bound is given  in Fig.~\ref{CMS}.

\section{Predictions for the Higgs couplings  from BSM solutions to the hierarchy problem}

The simplicity of the SM Higgs-mechanism is at odds with its quantum stability.
The fact that the Higgs is a scalar, a spin zero state, makes it difficult to keep it light ($m_h\ll M_P$).
This problematic  can be easily understood just by looking at  the  degrees of freedom (DOF) of a  massless and massive  state of spin 0, and compare them with those of a state of spin 1/2, 1, or higher.
Indeed, a massless vector, as the photon, has two polarizations (2 DOF), while a massive vector has 3 polarizations. The $2\not= 3$ guarantees that a massless vector can never get a mass by  continuous variations of parameters (or quantum fluctuations); only a discrete change in the theory, increasing the DOF, can make vector massive.
Similarly for fermions, we have that a charged massless fermion  has 2 DOF, while a massive one has the double (left- and right-handed states), and therefore, for the same reason,  massless fermions are  safe from getting masses under fluctuations.
\footnote{If a fermion has no charge, it can get a  Majorana-type mass without increasing the DOF, as probably is the case for the SM neutrinos. For this reason, to keep naturally massless fermions,  we must assume that the fermion has some type of  charge.}
Now, massless scalars have the same DOF as massive scalars: 1 DOF for neutral ones.
Even if we start with a massless scalar at tree-level, it is  not guaranteed that quantum corrections will not give it a mass.

A possible solution to keep the Higgs stable from getting a large mass
is to upgrade  the SM to include a symmetry  relating the Higgs, a scalar, to a fermion whose mass can be stable,
as we explained above. This is the case of  supersymmetry.
An alternative option is to assume that the Higgs is not an elementary state but a state made of elementary fermions, as pions in QCD. In this case, the Higgs  arises as a  composite state from a new strong-sector at the TeV.
It is interesting to point out that both scenarios predicted a light Higgs. While in  minimal supersymmetric versions of the SM 
(MSSM)  the lightest-Higgs mass was expected to be in the range $m_h\lesssim 135$ GeV \cite{Djouadi:2005gj},  
  minimal versions of  composite Higgs (MCHM)   predicted
  $115\ \text{GeV}\lesssim m_h\lesssim 185$ GeV \cite{Contino:2006qr}.
The connection between the Higgs mass and the mass spectrum of resonances is 
of crucial phenomenological interest, since allows 
to obtain predictions,
from  the present experimental value  $m_h\simeq 125$ GeV,
 for the heavy spectrum, either stops  for the MSSM \cite{Hall:2011aa}
or fermionic resonances for the MCHM \cite{CHconnection,Pomarol:2012qf}.

In the following,  we will centre in the predictions of  these models to  Higgs couplings.
As we emphasized in the introduction, the Higgs is usually the SM particle whose couplings are most sensitive to  BSM corrections.
Indeed, as we will see below,  in supersymmetric theories  Higgs  couplings can be affected at tree-level \cite{Arvanitaki:2011ck}, while other SM couplings are  affected at the loop level. 
Similarly, in strongly-interacting theories in which the Higgs is composite, effects on Higgs couplings can be enhanced by a factor $g_*^2$  \cite{Giudice:2007fh}, that can be as large as $\sim 16 \pi^2$, with respect to effects in other  couplings.
 It is also important to remark that in  BSMs trying to solve the hierarchy problem
the   main BSM effects in Higgs physics  
are captured  by the  primary Higgs couplings, as contributions to  non-primary Higgs couplings 
are usually negligible. This shows once more the importance of the  primaries.

\begin{figure}
\centering\includegraphics[width=.3\linewidth]{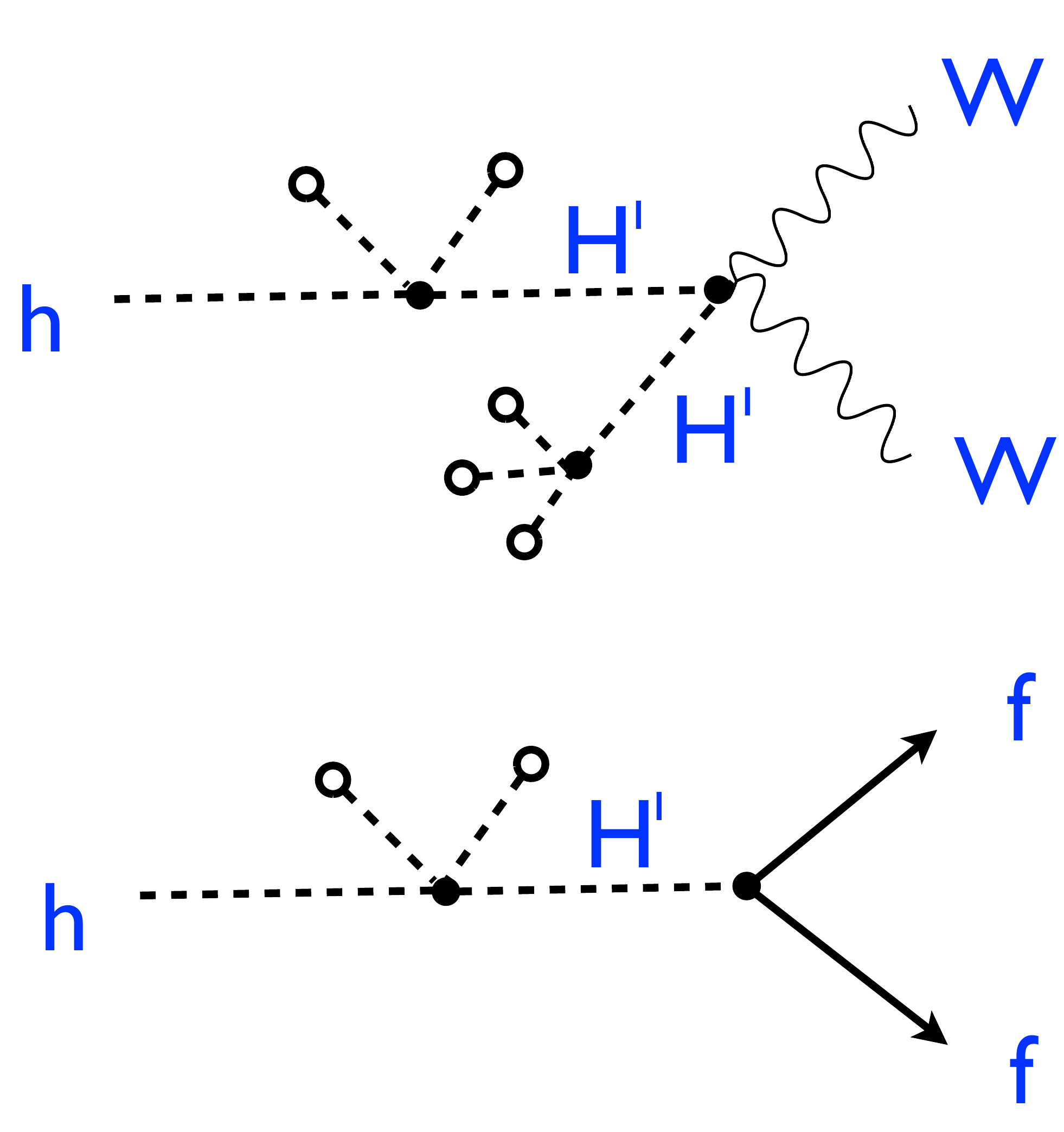}
\caption{Feynman diagrams contributing to  $g^h_{VV}$ and $g^h_{ff}$ from integrating the heavy MSSM Higgs $H'$.}
\label{MSSMcorrections}
\end{figure}

\begin{figure}
\centering
\includegraphics[width=0.47\textwidth]{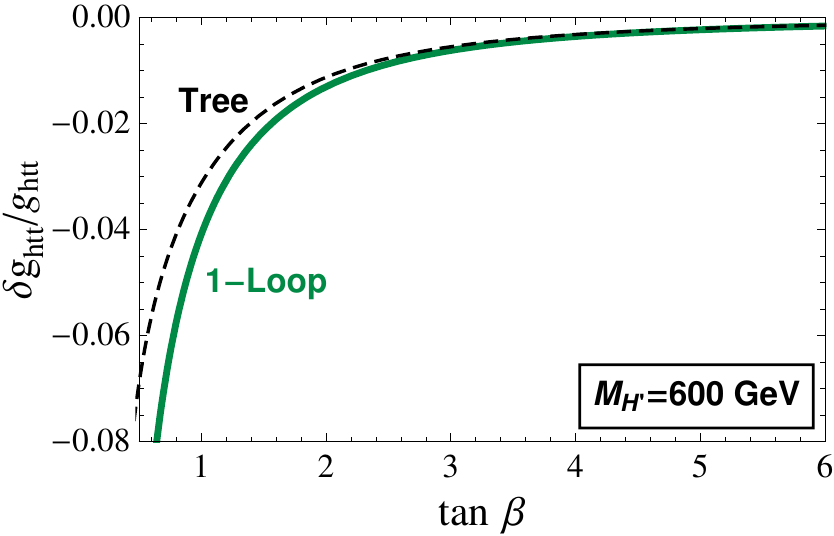}    \qquad
\includegraphics[width=0.46\textwidth]{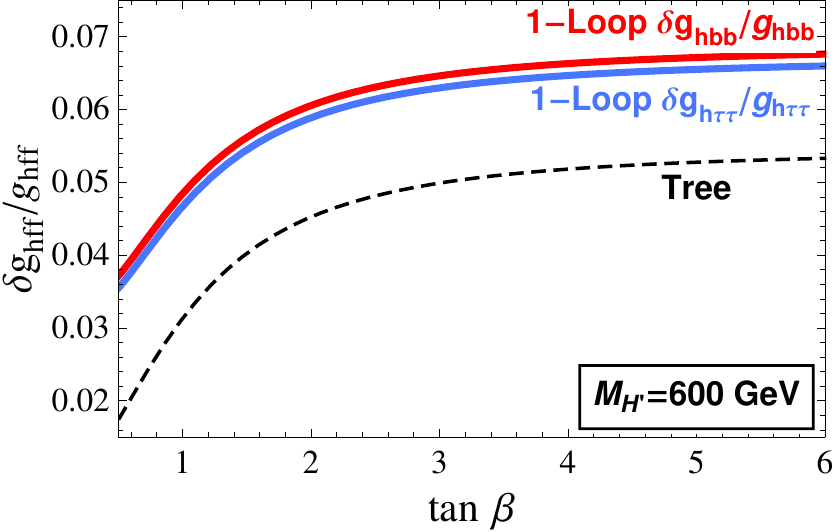}
\caption{Relative modifications of the Higgs couplings to fermions  with respect to their SM values
at tree-level (dashed line), and after including RGE effects from $\Lambda$ to the electroweak scale (solid lines), 
as a function of $\tan\beta$ in a MSSM scenario with $\Lambda=M_{H'}=600$ GeV and unmixed stops
heavy enough to reproduce $m_h=125$ GeV.  Left plot: Higgs coupling to tops.  Right plot:  Higgs coupling to bottoms (upper solid line) and taus (lower solid line)     \cite{Elias-Miro:2013mua}.}
\label{EEMP}
\end{figure}

\subsection{The Minmal Supersymmetric SM (MSSM)}

We will work in the  limit in which the supersymmetric spectrum is heavier than $m_h$.
This covers {\it most} of the parameter space of the MSSM, after  LHC  searches have 
pushed the  superpartner masses towards the TeV regime, and none deviation from the SM has been observed.
Also to accommodate $m_h\simeq 125$ GeV  requires large stop masses  in the MSSM \cite{Hall:2011aa}.

\begin{figure}
\centering\includegraphics[width=.55\linewidth]{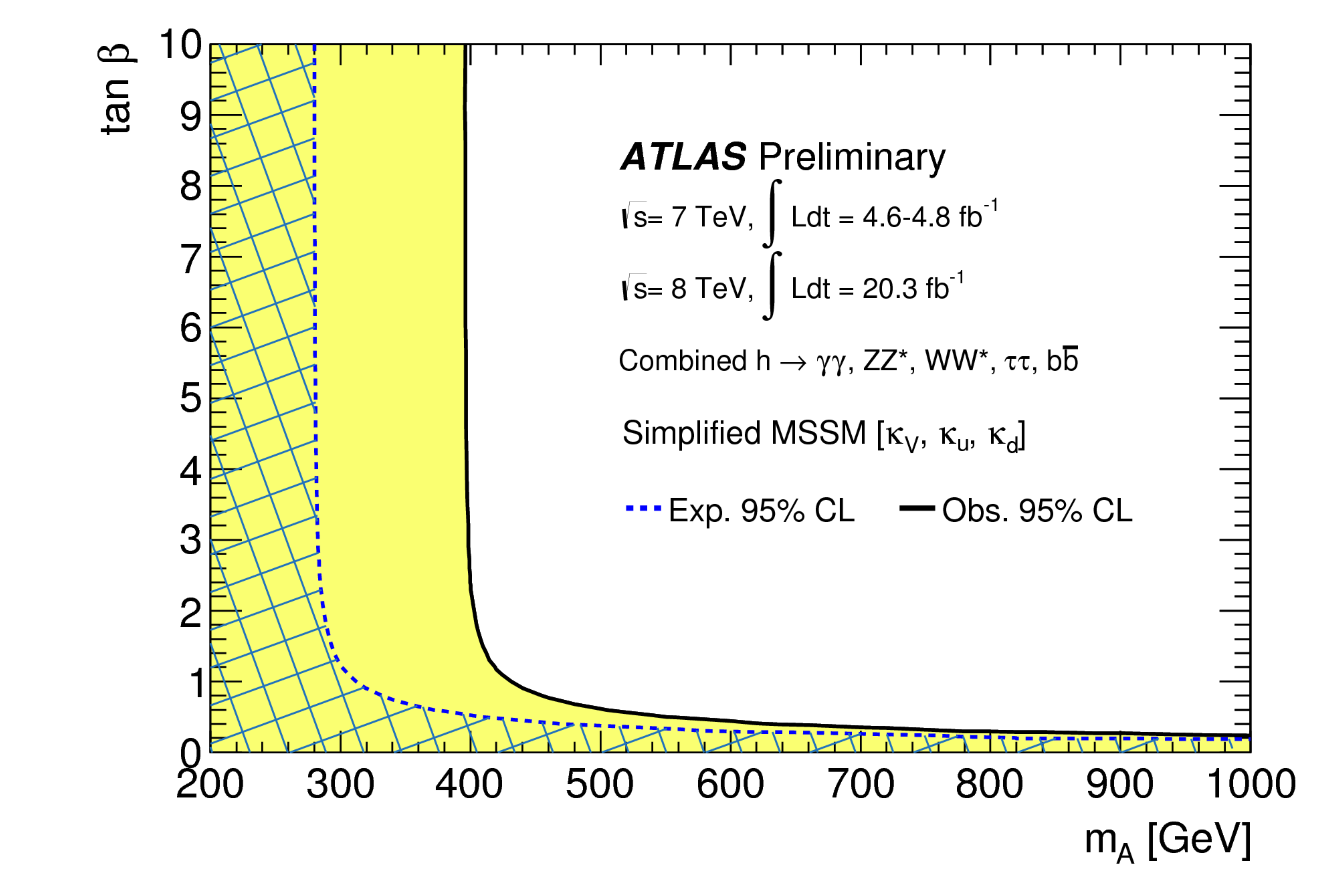}
\caption{Regions of the $m_A-\tan\beta$  plane excluded by Higgs physics in a  MSSM with heavy partners   \cite{ATLAS14-10}.}
\label{MSSM}
\end{figure}

The only tree-level corrections to the lightest-Higgs couplings  come from the extra heavy Higgs doublet of the MSSM $H'$.
This is due to the $R$-parity of the MSSM that only allow  $R$-even field   tree-level corrections.
At order $v^2/M_{H'}^2$ ({\it i.e.}, keeping only  $1/\Lambda^2$-suppressed effects where now $\Lambda=M_{H'}$), 
only the Higgs couplings to fermions are  affected, 
since corrections to   $hVV$  appear at order 
$v^4/M_{H'}^4$  as can be easily understood from Feynman diagrams --see Fig.~\ref{MSSMcorrections}.
 Deviations from the SM values for the $hff$  couplings, 
including also one-loop RGE effects coming from the top, are given by
\cite{Elias-Miro:2013mua}
\bea
\frac{\delta g^h_{tt}}{g^{h\, \rm SM}_{tt}}&=&\frac{v^2}{M^2_{H'}}\left(  \frac{\lambda^{\prime}}{t_\beta}\left[1-\frac{21y^2_t }{16\pi^2}\log\frac{M_{H'}}{m_h}\right]
+\frac{3y^4_t}{4\pi^2 t^2_\beta}\log\frac{M_{H'}}{m_h}\right)\nonumber \ ,\\
\frac{\delta g^h_{bb}}{g^{h\, \rm SM}_{bb}}&=& - \frac{v^2}{M^2_{H'}}\left(\lambda^{\prime }t_\beta\left[1- \frac{y^2_t}{2\pi^2} \log\frac{M_{H'}}{m_h}\right]+\frac{y^2_t}{16\pi^2}\left[5\frac{\lambda^{\prime }}{t_\beta}
- 14 y_t^2\right]\log\frac{M_{H'}}{m_h}\right) \ ,\nonumber\\
\frac{\delta g^h_{\tau\tau}}{g^{h\, \rm SM}_{\tau\tau}} &=& -\frac{v^2}{M^2_{H'}}\left( \lambda^{\prime }t_\beta\left[1 - \frac{3y_t^2}{8 \pi^2}   \log\frac{M_{H'}}{m_h} \right] + \frac{ 3y_t^2}{8 \pi^2} \left[\frac{\lambda^{\prime}}{t_\beta}- 2 y_t^2  \right] \log\frac{M_{H'}}{m_h}\right)
\ ,\label{ghff}
\eea
with $t_\beta\equiv \tan\beta$ and 
\footnote{In \eq{lp} we are also including RGE effects from $M_{\tilde t}$ to $M_{H'}$ proportional to the top-Yukawa $y_t$.}
\be
\lambda'=\frac{1}{8} (g^2+{g'}^2)\sin 4\beta-
\frac{3y_t^4}{8\pi^2t_\beta}\log\frac{M_{\tilde t}^2}{M_{H'}^2}\, ,
\label{lp}
\ee
where  $M_{\tilde t}$ is the value of the stop masses  taking, for simplicity, zero stop left-right mixing.
To illustrate the impact of these corrections, let us  take $M_{\tilde t}$ large enough to get $m_h\simeq 125$ GeV through the well-known loop
corrections to the Higgs quartic coupling, which at one-loop and neglecting stop left-right mixings read:
$\lambda=\frac{1}{8} (g^2+{g'}^2)\cos^22\beta+
\frac{3y_t^4}{16\pi^2}\log\frac{M_{\tilde t}^2}{M_t^2}$.
This gives the value of $\lambda'$ as a function of $t_\beta$ and $M_{H'}$  that we can then plug in Eq.~(\ref{ghff})
to obtain the  RGE-improved corrections for  $g^h_{ff}$  induced by integrating out the heavy Higgs. 
The results are shown in Fig.~\ref{EEMP}.  The experimental
bounds on the Higgs couplings can be translated into a bound on $M_{H'}$ as a function of $t_\beta$.
This is given in Fig.~\ref{MSSM}
where $m_A$ is the mass of the heavy MSSM CP-odd scalar that, at the order  we are working at, is equal to  $M_{H'}$.
\footnote{Mass splittings among  the heavy Higgs-doublet components  are $O(v^2/M^2_{H'})$, 
and then their effects are  of  higher-order in our expansion.}
Stop left-right  mixing effects or  extra $D$-term effects can be easily included along the lines of \cite{Gupta:2012fy}.
Corrections to $g_{3h}$ can also arise at order $O(v^2/M^2_{H'})$, but we already said that this coupling is difficult to 
measure as it requires double Higgs production.

\subsection{The Minimal Composite Higgs Model (MCHM)}

\begin{figure}
\centering\includegraphics[width=.65\linewidth]{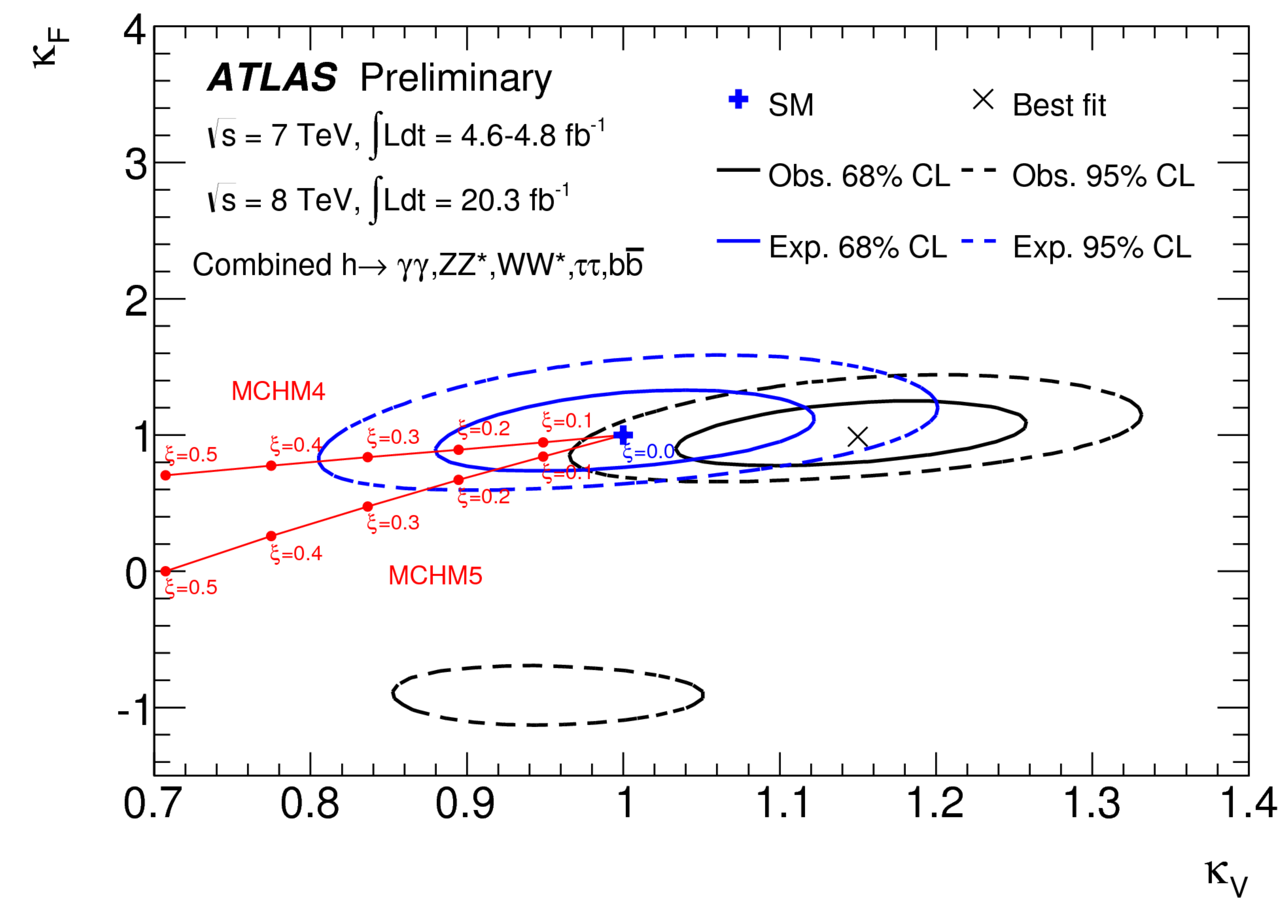}
\caption{Two-dimesional fit of  the Higgs couplings $\kappa_V\equiv g^h_{VV}/g^{h\,\rm SM}_{VV}$ and 
$\kappa_F\equiv g^h_{ff}/g^{h\,\rm SM}_{ff}$ and predictions from the MCHM4 and MCHM5 as a function of $\xi\equiv v^2/f^2$ \cite{ATLAS14-10}.}
\label{MCHM}
\end{figure}

For models in which  the Higgs is a pseudo Goldstone boson (PGB) arising from a new strong-sector at the TeV 
\cite{reviews}, 
similar to a pion in QCD, the Higgs couplings must depart from their SM value.
This was studied in generality in \cite{Giudice:2007fh}. The main effects are expected to arise
in the Higgs coupling to $Z/W$ and fermions.
The minimal model is the MCHM \cite{Agashe:2004rs}, where the global symmetry-breaking pattern is $SO(5)\to SO(4)$
with an "order parameter" $f$, that give the following   predictions \cite{Giudice:2007fh}:
\bea
\frac{g^h_{VV}}{g^{h\, \rm SM}_{VV}}&=&\sqrt{1-\frac{v^2}{f^2}}\, ,\nn\\
\frac{g^h_{ff}}{g^{h\, \rm SM}_{ff}}&=&\frac{1-(1+n)v^2/f^2}{\sqrt{1-{v^2}/{f^2}}}\, ,
\label{MCHMcouplings}
\eea
where $n=0,1,2,...$ depends on how fermions are implemented in the model. 
In particular, for  the MCHM4 (MCHM5) we  have $n=0\, (1)$ \cite{Pomarol:2012qf}. 
From the minimization of the Higgs potential, we expect $f\gtrsim v$ \cite{Giudice:2007fh},
but constraints from the $\widehat S$ parameter give $v^2/f^2\lesssim 0.1$ \cite{reviews}.
The Higgs coupling predictions of the MCHM are shown in  Fig.~\ref{MCHM} and compared with a fit of the ATLAS data.
The fact that   the experimental data does not favour smaller   Higgs couplings than those of the SM, as 
predicted from \eq{MCHMcouplings},  implies that we can derive an upper  bound on $\xi\equiv v^2/f^2$,
and consequently  on the composite scale, $\Lambda\simeq g_* f$, where $g_*$ is here the coupling among the resonances of the strong sector, expected to be in the range, $1\ll g_*\lesssim 4\pi$.
ATLAS  \cite{ATLAS14-10} gives the observed (expected) $95\%$ CL upper limit of $\xi< 0.12\ (0.29)$
for the MCHM4 and $\xi< 0.15\ (0.20)$ for the  MCHM5 that start being  as competitive as the ones 
coming from LEP \cite{reviews}.

Contributions to $\kappa_{\gamma\gamma}$  and $\kappa_{GG}$
are suppressed in the MCHM due to the PGB nature of the Higgs \cite{Giudice:2007fh}.
Nevertheless, this suppression is not present in  $\kappa_{Z\gamma}$ that can receive significant contributions
 \cite{Azatov:2013ura} that could be even larger than those of the SM, providing    a strong motivation for  searching for $h\to Z\gamma$.

\section{Conclusions}

With the Higgs discovery, the full SM has been  experimentally   established.
Nevertheless, the presence of the Higgs, a zero-spin state,  demands 
new physics at the TeV to make the SM a natural theory.
The Higgs is the most sensitive SM particle to new physics,
and for this reason an accurate measurement of its couplings
provides    an excellent  way to   indirectly discover new phenomena.

At the LHC (and  in  future colliders)
we can have   access to a large variety of  Higgs couplings. 
 We have argued that the most relevant Higgs couplings  are the primary ones, given in \eq{prim} for CP-conservation.
These couplings probe new directions in the parameter space of BSMs.
We have showed the predictions for two of the most well-motivated BSMs,  the MSSM and the MCHM.
These analysis can be extended to other BSMs, such as the  non-minimal MSSM (NMSSM),
or other possibilities for composite Higgs,  for example those in which the Higgs is  lighter than the composite scale $\Lambda$ not because of its PGB nature, as in the MCHM, but due to  an "accidental"  supersymmetry (SUSY Composite Higgs) or  scale symmetry (Higgs as a dilaton) \cite{reviews}. Supersymmetry can also allow for   a partly-composite Higgs where the  TeV strong-sector could also break the electroweak symmetry  (bosonic TC) \cite{SSSB}.
A brief summary of the largest effects in the primary Higgs coupling 
 arising from these scenarios   is  giving in Table~1.
If in the future departures from the SM  Higgs couplings are observed,   the  analysis of  the pattern  of these deviations will  be extremely useful to  discriminate between different BSM scenarios.

{\renewcommand{\arraystretch}{1.4} 
\begin{table}[htdp]
\begin{center}\begin{tabular}{|c|c|c|c|c|c|c|}
\hline
&  $g^h_{ff}$ & $g^h_{VV}$ & $\kappa_{GG}$ & $\kappa_{\gamma\gamma}$ & $\kappa_{Z\gamma}$ & $g_{3h}$
 \\\hline
 \hline MSSM & \checkmark &  &  &  &  &\checkmark
 \\\hline NMSSM 
& \checkmark & \checkmark &   \checkmark&  \checkmark & \checkmark &\checkmark
 \\\hline MCHM & \checkmark & \checkmark &  &  & \checkmark &\checkmark
 \\\hline SUSY Composite Higgs& \checkmark & \checkmark &  &  &  &\checkmark
 \\\hline Higgs as a Dilaton  & &  &   \checkmark&  \checkmark & \checkmark &\checkmark
  \\\hline Partly-Composite Higgs  &  &  &   \checkmark&  \checkmark & \checkmark &\checkmark
 \\\hline Bosonic TC &  &  &  &  &  & \checkmark 
\\\hline
 \end{tabular} \caption{Largest  contributions  to Higgs couplings (relative to the SM one) expected from different  BSM scenarios.}
\end{center}
\label{defaulttable}
\end{table}
}

\vskip.5cm
\section*{Acknowledgements}
I  am  very thankful to  Christophe Grojean, Eduard Masso and Francesco Riva  for 
useful comments on the manuscript.
This work has been partly supported by the grants FPA2011-25948,  2009SGR894
and the Catalan ICREA Academia Program.

\end{document}